\documentclass[twocolumn,tighten]{aastex63}
\usepackage{amsmath}
\usepackage{threeparttable}
\usepackage{float}
\usepackage{graphicx}
\usepackage{CJK}

\newcommand{\package}[1]{\tt{#1}}

\shorttitle{Reconstructing GSE}
\shortauthors{Naidu et al.}

\begin{document}
\begin{CJK*}{UTF8}{gbsn}

\title{Reconstructing the Last Major Merger of the Milky Way with the H3 Survey}

\correspondingauthor{Rohan P. Naidu}
\email{rohan.naidu@cfa.harvard.edu}
\author[0000-0003-3997-5705]{Rohan P. Naidu}
\affiliation{Center for Astrophysics $|$ Harvard \& Smithsonian, 60 Garden Street, Cambridge, MA 02138, USA}
\author[0000-0002-1590-8551]{Charlie Conroy}
\affiliation{Center for Astrophysics $|$ Harvard \& Smithsonian, 60 Garden Street, Cambridge, MA 02138, USA}
\author[0000-0002-7846-9787]{Ana Bonaca}
\affiliation{Center for Astrophysics $|$ Harvard \& Smithsonian, 60 Garden Street, Cambridge, MA 02138, USA}
\author[0000-0002-5177-727X]{Dennis Zaritsky}
\affiliation{Steward Observatory, University of Arizona, 933 North Cherry Avenue, Tucson, AZ 85721-0065, USA}
\author[0000-0001-6260-9709]{Rainer Weinberger}
\affiliation{Center for Astrophysics $|$ Harvard \& Smithsonian, 60 Garden Street, Cambridge, MA 02138, USA}
\author[0000-0001-5082-9536]{Yuan-Sen Ting (丁源森)}
\altaffiliation{Hubble Fellow}
\affiliation{Institute for Advanced Study, Princeton, NJ 08540, USA}
\affiliation{Department of Astrophysical Sciences, Princeton University, Princeton, NJ 08544, USA}
\affiliation{Observatories of the Carnegie Institution of Washington, 813 Santa Barbara Street, Pasadena, CA 91101, USA}
\affiliation{Research School of Astronomy and Astrophysics, Mount Stromlo Observatory, Cotter Road, Weston Creek, ACT 2611, Canberra, Australia}
\author[0000-0003-2352-3202]{Nelson Caldwell}
\affiliation{Center for Astrophysics $|$ Harvard \& Smithsonian, 60 Garden Street, Cambridge, MA 02138, USA}
\author[0000-0002-8224-4505]{Sandro Tacchella}
\affiliation{Center for Astrophysics $|$ Harvard \& Smithsonian, 60 Garden Street, Cambridge, MA 02138, USA}
\author[0000-0002-6800-5778]{Jiwon Jesse Han}
\affiliation{Center for Astrophysics $|$ Harvard \& Smithsonian, 60 Garden Street, Cambridge, MA 02138, USA}
\author[0000-0003-2573-9832]{Joshua S. Speagle}
\affiliation{University of Toronto, Department of Statistical Sciences, Toronto, M5S 3G3, Canada}
\affiliation{University of Toronto, David A. Dunlap Department of Astronomy \& Astrophysics, Toronto, M5S 3H4, Canada}
\affiliation{Dunlap Institute for Astronomy \& Astrophysics, Toronto, M5S 3H4, Canada}
\author[0000-0002-1617-8917]{Phillip A. Cargile}
\affiliation{Center for Astrophysics $|$ Harvard \& Smithsonian, 60 Garden Street, Cambridge, MA 02138, USA}

\begin{abstract}
Several lines of evidence suggest that the Milky Way underwent a major merger at $z\sim2$ with a galaxy known as \textit{Gaia}-Sausage-Enceladus (GSE).  Here we use H3 Survey data to argue that GSE entered the Galaxy on a retrograde orbit based on a population of highly retrograde stars with chemistry similar to the largely radial GSE debris. We present the first tailored, high-resolution ($10^{4}-10^{5} M_{\rm{\odot}}$) N-body simulations of the merger. From a grid of $\approx500$ simulations we find a GSE with $M_{\star}=5\times10^{8}\ M_{\rm{\odot}}, M_{\rm{DM}}=2\times10^{11} M_{\rm{\odot}}$ (a 2.5:1 total mass merger) accreted on an inclined ($15^{\circ}$), intermediate circularity (0.5) orbit best matches the H3 data. This simulation shows the retrograde GSE stars are stripped from its outer disk early in the merger before the orbit loses significant angular momentum. Despite being selected purely on angular momenta and radial distributions, this simulation reproduces and explains the following empirical phenomena: (i) the elongated, triaxial shape of the inner halo (axis ratios $10:7.9:4.5$), whose major axis is at $\approx35^{\circ}$ to the plane and connects GSE's apocenters, (ii) the Hercules-Aquila Cloud \& the Virgo Overdensity, which arise due to apocenter pile-up on either end of the major axis, (iii) the 2 Gyr lag between the quenching of GSE and the truncation of the age distribution of the in-situ halo, which tracks the 2 Gyr gap between the first and final GSE pericenters. We make the following predictions: (i) the inner halo has a ``double-break" density profile with breaks at both $\approx15-18$ kpc and $30$ kpc, coincident with the GSE apocenters, (ii) the outer halo is highly structured, with retrograde streams containing $\approx10\%$ of GSE stars awaiting discovery at $>30$ kpc. The retrograde (radial) GSE debris originates from its outer (inner) disk -- exploiting this trend we reconstruct the stellar metallicity gradient in a $z\approx2$ star-forming galaxy ($-0.04\pm0.01$ dex $r_{\rm{50}}^{-1}$). These simulations imply the GSE merger delivered $\approx20\%$ of the Milky Way's present-day dark matter and $\approx50\%$ of its stellar halo.
\end{abstract}


\section{Introduction}
\label{sec:introduction}

A hallmark feature of $\Lambda$CDM cosmology is hierarchical assembly, in which galaxies continually assimilate smaller systems \citep[e.g.,][]{White91}. Nowhere in the Universe do we have a clearer view of this hierarchical build-up than in the stellar halo of the Milky Way (MW). At this very moment the Sagittarius dwarf galaxy is being tidally disrupted \citep[e.g.,][]{Ibata94}, the Magellanic Clouds are on first infall \citep[e.g.,][]{Besla07}, and dozens of globular cluster streams encircle the Galaxy \citep[e.g.,][]{Bonaca20b}.

While these ongoing mergers are apparent on the sky, the record of even more mixed, ancient mergers can be extracted from the stellar halo. Due to the long relaxation time in the halo, stars that were accreted as part of the same galaxy can be connected through their shared integrals of motion (e.g., angular momenta, energies) even several Gyrs after their arrival \citep[e.g.,][]{Helmi00,Font11,Simpson19}. We are also aided by the shared chemical abundance patterns expected of stars born in the same system \citep[e.g.,][]{Freeman02,Venn04,Lee15}. Integrals of motion and chemical information have recently been obtained for millions of stars in the solar neighborhood thanks to the \textit{Gaia} mission \citep{gaia} and  stellar spectroscopic surveys such as APOGEE \citep{APOGEE}, RAVE \citep{RAVE}, SEGUE \citep{SEGUE}, LAMOST \citep{LAMOST}, GALAH \citep{GALAH}, and H3 \citep[][]{Conroy19}. These data have allowed us to piece together the history of the Galaxy in unprecedented detail.

A single dwarf galaxy that merged with the Milky Way at $z\approx2$ -- \textit{Gaia} Sausage Enceladus (GSE) -- constitutes the bulk of the inner halo \citep[e.g.,][]{Belokurov18,Helmi18,Naidu20}. The lines of evidence for this accretion event are numerous and compelling. Kinematics of halo stars show a preponderance of eccentric, radial orbits \citep[e.g.,][]{Eggen62,Chiba00,Koppelman18,Mackereth19,Carollo20,Yuan20, Limberg21} exactly as expected for debris from a major merger that is radialized due to dynamical friction \citep[e.g.,][]{Amorisco17}. The ages and abundances of these eccentric stars point to the same, ancient (${\gtrsim}8-10$ Gyr old) progenitor \citep[e.g.,][]{Haywood18,Gallart19,Conroy19b,Bonaca20,Das20,Feuillet20,Gudin21}. A large number of MW globular clusters ($\approx20-30$) are eccentric and clustered in the age-metallicity plane, suggesting they accompanied GSE to the MW \citep[e.g.,][]{Myeong18,Massari19,Kruijssen19,Forbes20}. A break in the halo density and anisotropy profiles at $\approx25-30$ kpc has been associated with an apocenter in the GSE orbit \citep[e.g.,][]{Deason18,Lancaster19,Bird19,Iorio21}. These observations are supported by cosmological simulations that show the inner halos of MW-like galaxies are often built out of a handful of massive progenitors, and that a large fraction of debris from these mergers often ends up on eccentric orbits \citep[e.g.,][]{Deason15,Fattahi19,Grand20,Santistevan20}.

A fundamental open question is the configuration of the merger -- did GSE collide with our Galaxy head-on, or was it on an initially circular orbit that decayed? Recovering the configuration hinges on whether GSE debris today is purely radial or if it extends to orbits with significant angular momentum \citep[e.g.,][]{Evans20,Helmi20,Koppelman20b}. The configuration informs a variety of issues e.g., the expected velocity and spatial distributions of the dark matter (DM) that arrived with GSE ($\approx20\%$ of the MW's DM, see \S\ref{sec:fiducial}). The velocity distribution modulates the expected signal in DM detection experiments, and a significant non-radial component could influence efforts seeking directional signatures \citep[e.g.,][]{OHare18,OHare20,Evans19,Vahsen20}. Similarly, a merger configuration resulting in a non-planar mass distribution that breaks axisymmetry would have important implications for the MW potential, particularly in the inner halo ($<30$ kpc) which is essentially entirely comprised of GSE \citep[][]{Naidu20}.

The debate around the radial or retrograde nature of GSE has largely relied on local halo samples that are limited to a few kpc from the Sun. However, the first stars stripped from GSE, which contain the most information about its initial orbit, are likely at larger distances and higher energies than stars that pass through the solar neighborhood. Capturing this early debris, which retains the most pristine memory of the merger configuration, requires forging beyond the local halo. 

The H3 Stellar Spectroscopic Survey \citep{Conroy19} is designed to study the distant halo. Combined with \textit{Gaia}, H3 is measuring full 6D phase-space coordinates and chemical abundances for $\approx200,000$ stars at $r_{\rm{gal}}\approx3-100$ kpc. Using these data, \citet{Naidu20} presented a comprehensive inventory of structure in the halo out to $50$ kpc, including the largest sample of GSE (N=2684) stars with integrals of motion and abundances from high-resolution spectroscopy. This sample is unique in encompassing the farthest reaches of the merger, and in being largely metallicity-unbiased (unlike e.g., RR Lyrae or BHB or color-selected samples). In this work, we build on \citet{Naidu20} to explore the retrograde halo with a view to chart the full extent of GSE.

Tailored simulations of the other significant MW mergers -- Sagittarius \citep[e.g.,][]{LM10,Dierickx17,Laporte18} and the Magellanic Clouds \citep[e.g.,][]{Besla10,Garavito-Camargo19,Vasiliev20} -- have proven crucial in interpreting phenomena across the Galaxy such as the phase-space spiral \citep[e.g.,][]{Antoja18,Bland-Hawthorn19} and the reflex motion of the outer halo \citep[e.g.,][]{Petersen20,Erkal20}. However, the retrograde or radial nature of GSE is still unclear, and most existing constraints on the merger are derived from the local halo, inhibiting the production of a high fidelity model. Consequently, GSE has been studied largely qualitatively via analogs in cosmological simulations \citep[][]{Bignone19,Elias20}, Milky Way zooms \citep[][]{Fattahi19,Grand20}, and existing merger simulations \citep[][]{Helmi18,Koppelman20b}. These studies have been immensely successful in demonstrating how a major merger can produce eccentric debris and reshape the early MW disk. Equipped with constraints from the H3 Survey we are well-positioned to build on these results and produce a tailored model for the merger as has been done for Sagittarius and the LMC.

A plan for the paper follows. In \S\ref{sec:datamain} we argue that a subset of the retrograde halo stars are associated with the GSE merger. In \S\ref{sec:constraints} we summarize existing observational constraints on the merger. \S\ref{sec:sims} describes the numerical simulations. \S\ref{sec:results} is based on the fiducial simulation -- here we interpret the origin of GSE's highly retrograde debris (\S\ref{sec:arjuna_origin}), the shape of the inner halo (\S\ref{sec:shape}), the all-sky distribution of GSE debris (\S\ref{sec:outerhalo}), the inner halo density profile (\S\ref{sec:profile}), the timeline of the GSE merger (\S\ref{sec:timing}), the net rotation of GSE (\S\ref{sec:netrot}), and the relationship between GSE and other retrograde accreted galaxies (\S\ref{sec:sequoia}). In \S\ref{sec:fehgrad} we use our fiducial simulation to reconstruct the stellar metallicity gradient measurement in a $z\approx2$ star-forming galaxy (GSE). A summary follows in \S\ref{sec:summary}.

We adopt a \citet{Planck18} cosmology. To describe central values of distributions we generally report the median, along with 16th and 84th percentiles. We use $r_{\rm{gal}}$ to denote 3D Galactocentric distance, $X_{\rm{gal}}, Y_{\rm{gal}}, Z_{\rm{gal}}$ to denote Galactocentric Cartesian distances, and $d_{\rm{helio}}$ to refer to 3D heliocentric distance. We use $V_{r}$,  $V_{\rm{\phi}}$,  $V_{\rm{\theta}}$ for velocities in a right-handed spherical coordinate system with origin at the Galactic center. Prograde stars have negative $V_{\rm{\phi}}$ and $L_{\rm{z}}$. Unless mentioned otherwise, total orbital energy ($E_{\rm{tot}}$) is always reported in units of $10^{5}\ \rm{km^{2}\ s^{-2}}$ and angular momenta ($L_{\rm{x}}$, $L_{\rm{y}}$, $L_{\rm{z}}$) in units of $10^{3}\ \rm{kpc}\ \rm{km\ s^{-1}}$. These quantities are always computed in a Galactocentric frame tied to the center of the Milky Way both in the data and the simulations.

\begin{figure}[t]
\centering
\includegraphics[width=0.9\linewidth]{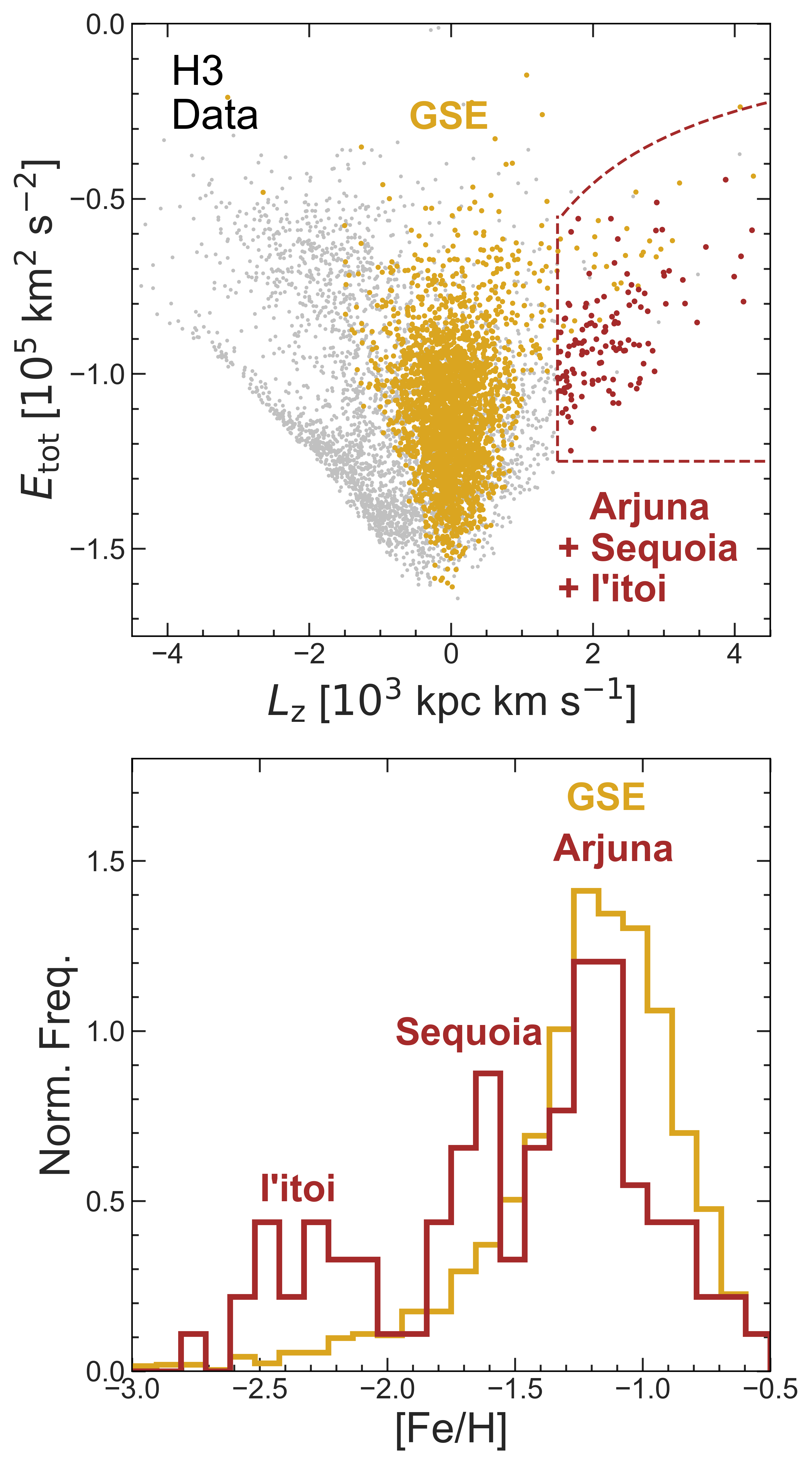}
\caption{Arjuna as the highly retrograde debris of GSE. \textbf{Top:} $E-L_{\rm{z}}$ diagram plotting the total energies of the H3 giants (gray) against the $z$-component of their angular momenta. GSE, defined to lie at eccentricities $>0.7$ is shown in gold, whereas the high-energy retrograde halo defined by the dashed lines is shown in brown. \textbf{Bottom:} Metallicity distribution function (MDF) of the GSE stars compared with retrograde stars. The retrograde MDF shows three populations -- I'itoi at [Fe/H]$<-2$, Sequoia at [Fe/H]$\approx-1.6$, and Arjuna at [Fe/H]$\approx-1.2$. The Arjuna MDF closely tracks the GSE MDF.}
\label{fig:data}
\end{figure}

\section{Revealing the full extent of GSE}
\label{sec:datamain}
\subsection{Data: The H3 Survey}
\label{sec:data}

The H3 Survey \citep[][]{Conroy19} is a high-latitude ($|b|>30^{\circ}$), high-resolution ($R=$32,000) spectroscopic survey of the distant ($d_{\rm{helio}}\approx2-100$ kpc) Galaxy. Targets are selected purely on their \textit{Gaia} parallax ($\pi<0.4-0.5$ mas, evolving with \textit{Gaia} data releases), brightness ($15<r<18$), and observability (Dec.$>-20^{\circ}$) from the 6.5m MMT in Arizona, USA. H3 is measuring radial velocities precise to $\lesssim$1 km $\rm{s^{-1}}$, [Fe/H] and [$\alpha$/Fe] abundances precise to $\lesssim$0.1 dex, and spectrophotometric distances precise to $\lesssim$10$\%$  (see \citealt{Cargile20} for details of the stellar parameter pipeline). Combined with \textit{Gaia} proper motions (SNR$>$3 for $>$90$\%$ of the sample), H3 thus provides the full 6D phase-space and 2D chemical-space for all stars in the sample.

\citet[][]{Naidu20} used the sample of H3 giants (N=5684, $|b|{>}40^{\circ}$, $d_{\rm{helio}}=3-50$ kpc) to assign almost the entire distant Galaxy to various structures (summarized in their Table 1). These authors systematically identified debris from known accreted galaxies (e.g., Sgr, GSE, Sequoia), as well as new structures (e.g., I'itoi, Arjuna, Wukong). Pertinent to the matter at hand, they tagged $\approx3000$ stars as belonging to GSE or the high-energy retrograde halo that we focus on here. All quantities sourced from \citet[][]{Naidu20} (e.g., $L_{\rm{z}}$ and $r_{\rm{gal}}$ distributions) describe the $|b|>40^{\circ}$ Galaxy, and have been corrected for the H3 selection function -- in particular, for the survey magnitude limit and targeting algorithm (see their \S2.3). We make no corrections for the window function and compare models and data only within the survey footprint.

\subsection{Arjuna as the retrograde debris of GSE}
\label{sec:retroge}

After excluding Sgr, the high-$\alpha$ disk, and in-situ halo, \citet[][]{Naidu20} attributed stars on highly eccentric orbits ($e>0.7$) to GSE (N=2684). This is essentially the head-on ``\textit{Gaia}-Sausage" from \citet[][]{Belokurov18}. The resulting, well-sampled MDF is unimodal (median [Fe/H]$=-1.15$), consistent with a simple chemical evolution model \citep[``Best Accretion Model",][]{Lynden-Bell75,Kirby11}, and resembles the narrow MDFs of local dwarfs like Fornax and Leo I \citep[][]{Kirby13}.

The high-energy retrograde halo is defined in \citet[][]{Naidu20} as excluding GSE and by the following condition: $(\eta>0.15) \land (L_{\rm{z}}>0.7)
\land\ (E_{\rm{tot}}>-1.25)$, where $\eta$ is the orbital circularity computed as $L_{\rm{z}}/|L_{\rm{z,max}}(E_{\rm{tot}})|$, where $L_{\rm{z,max}}(E_{\rm{tot}})$ is the maximum $L_{\rm{z}}$ achievable for an orbit of energy $E_{\rm{tot}}$. This definition generously selects stars on retrograde orbits, and excludes the Thamnos structure \citep[][]{Koppelman19} at lower energy. In Figure \ref{fig:data} we further limit the high-energy retrograde halo to $L_{\rm{z}}>1.5$ to make it clear that the radial locus of stars typically associated with GSE (distributed around $L_{\rm{z}}=0$) is not responsible for the features discussed below. Three chemical populations comprise the high-energy, highly retrograde halo: Arjuna ([Fe/H]$\approx-1.2$), Sequoia ([Fe/H]$\approx-1.6$), and I'itoi ([Fe/H]$<-2$). We emphasize that these three chemical populations do not just occur along the margins of GSE in $E-L_{\rm{z}}$, but extend to highly retrograde orbits. The Sequoia MDF peaks exactly where other studies have found it to peak \citep[e.g.,][]{Myeong19,Matsuno19,Monty19} and I'itoi is a distinct metal-poor population. As foreshadowed in \citet[][]{Naidu20}, we argue here that the metal-rich Arjuna is the retrograde debris of GSE. 

The Arjuna MDF closely tracks the GSE MDF (bottom panel of Figure \ref{fig:data}), with a similar mode, similar mean metallicity, but fewer metal-rich stars. Not only this, the $\alpha$ abundances of GSE and Arjuna are virtually identical -- median [$\alpha$/Fe] of 0.21 and 0.24 respectively. Due to these similarities, we associate Arjuna with GSE. The Arjuna debris extends to very retrograde orbits ($L_{\rm{z}}\approx4$), is more distant (median $r_{\rm{gal}}\approx23$ kpc vs. 18 kpc for GSE) and less eccentric ($e=0.55$). In the sections that follow, we demonstrate through numerical simulations that Arjuna's properties are consistent with it being material from the outer regions of GSE. This material may have been stripped before the satellite was radialized, and so retains the high retrograde angular momentum of the early orbit of GSE. Further, since this material is shed during the early phase of the merger, it has a larger mean distance from the Galactic center compared to debris stripped at later times.

Before moving on, we briefly consider an alternative scenario: Arjuna as an [Fe/H]$=-1.2$ dwarf galaxy that despite having virtually identical abundances has nothing to do with GSE. It would be a significant coincidence for two distinct accreted dwarf galaxies to have mean [Fe/H] as well as mean [$\alpha$/Fe] within 0.05 dex. Further, the relative star-counts from H3 imply Arjuna is only $\approx5\%$ of the GSE stellar mass (i.e., $\approx10^{7} M_{\odot}$). This stellar mass and the measured metallicity ($-1.2$) together constrain the accretion epoch of the hypothetical Arjuna dwarf to be $z\approx0$ according to the redshift evolution of the mass-metallicity relation \citep{Kirby13, Ma16}. However, we would then expect the very recently accreted ($z\approx0$) Arjuna to be rather coherent on the sky (a la Sgr), but this is not the case. For these reasons we disfavor the interpretation of Arjuna as an unrelated dwarf galaxy.

A natural question is why the highly retrograde Arjuna, the most dominant component of the high-energy retrograde halo ($2\times$ as many stars as Sequoia) was not prominent in the local halo datasets (typically limited to $d_{\rm{helio}}\lesssim5$ kpc) used to study GSE \citep[e.g.,][]{Myeong19,Koppelman19,Helmi20}. From orbit integration we find that the Arjuna stars observed by H3 spend $\approx20\times$ less time in the solar neighborhood ($d_{\rm{helio}}<5$ kpc) than the local halo GSE samples used in these studies. The H3 Arjuna stars on average have larger apocenters, higher energies, and are at higher Galactic latitudes. The discovery of Arjuna underscores the value of surveying the distant halo.

\section{Summary of constraints on the GSE merger}
\label{sec:constraints}
Here we list measurements pertaining to the merger that we will use to guide our numerical experiments in \S\ref{sec:sims}.  While various datasets have been mined to shed light on GSE, we will constrain our simulations purely to measurements from the H3 Survey for consistency. Other measurements are used as independent cross-checks. The H3 constraints listed below apply to GSE as it appears within the survey footprint, and have been corrected for the selection function (\S\ref{sec:data}).

\begin{enumerate}

    \item \textbf{Existence of Arjuna}: While the bulk of GSE debris is on highly eccentric, radial orbits that appear as the ``sausage" overdensity centered at $V_{\rm{r}}\sim0$ in the $V_{\rm{r}}-V_{\rm{\phi}}$ plane, in this work we argue that the highly retrograde Arjuna also belongs to GSE. In particular, $\approx75\%$ of GSE debris is radial, with $|L_{\rm{z}}|<0.5$, while $\approx5\%$ extends to highly retrograde, high-energy orbits with $L_{\rm{z}}>1.5$.  See \S\ref{sec:retroge} for details.
    
    \item \textbf{Spatial distribution of GSE debris}: At $<50$ kpc, $\approx90\%$ of GSE debris is contained within $r_{\rm{gal}}\approx30$ kpc, $\approx60\%$ within $r_{\rm{gal}}\approx20$ kpc and $\approx10\%$ within $r_{\rm{gal}}\approx10$ kpc. Profiles of the halo using other datasets also show a break at $25-30$ kpc that has been associated with an apocenter of GSE \citep[e.g.,][]{Deason18,Lancaster19}. Further, the shape of the inner halo, which is dominated by GSE, has been measured by several authors \citep[e.g.,][]{Juric08,Xue15,Das16b}. We will use the recent all-sky \textit{Gaia} RR Lyrae constraints from \citet[][]{Iorio18,Iorio19} who found the inner halo defines a trixial ellipsoid (axis ratios $10:7.9:4.5$) as a  cross-check on the debris geometry.
    
    \item \textbf{Hercules-Aquila Cloud (HAC) and Virgo Overdensity (VOD)}: The HAC and VOD are large, diffuse stellar overdensities occurring on either side of the plane that have been known for more than a decade \citep[][]{Vivas01,Newberg02,Belokurov07,Juric08,Bonaca12}. Recently, thanks to \textit{Gaia}, both these structures have been linked to GSE based on the integrals of motion and eccentric orbits of their constituent stars (\citealt{Simion18,Simion19}, but see \citealt{Donlon19,Donlon20}). We will use the emergence of HAC and VOD-like structures at the appropriate locations as an independent cross-check on the models that best reproduce the H3 data.
    
    \item \textbf{Stellar mass of GSE}: Estimates of the stellar mass of GSE range from $\sim2-7\times10^{8} M_{\odot}$ and have been derived using the mass-metallicity relation assuming $z_{\rm{acc.}}\approx2$ ($\approx4-7\times10^{8} M_{\odot}$, \citealt[][]{Naidu20}), the age-metallicity \& dynamical clustering of accreted GSE GCs ($\approx2-4\times10^{8}\,M_{\odot}$, \citealt[][]{Kruijssen20}), counts of metal-poor ([Fe/H]$<-1$) eccentric ($e>0.7$) stars ($\approx2-5\times10^{8}\,M_{\odot}$, \citealt[][]{Mackereth20}), and from chemical evolution models ($\approx5-6\times10^{8}\,M_{\odot}$, \citealt[][]{Helmi18, Fernandez-Alvar18}).
    
    \item \textbf{Spatial extent of the in-situ halo}: A substantial fraction of the local kinematic halo is comprised of stars that have chemistry identical to the high-$\alpha$ disk, but that are on orbits with eccentricities higher than typical disk stars \citep[e.g.,][]{Nissen10, Bonaca17,Haywood18,Belokurov20,An21}. This ``in-situ halo"/``splash" is composed of stars kicked out of the primordial disk during the GSE merger, and its properties are sensitive to the mass of the MW and GSE at the time of the merger \citep[e.g.,][]{Fattahi19,Grand20}. \citet{Naidu20} chart the spatial extent of the in-situ halo (defined to have $e>0.5$ and high-$\alpha$ disk-like chemistry) and find $>90\%$ of it is confined to $r_{\rm{gal}}<20$ kpc and $|Z_{\rm{gal}}|<15$ kpc.
    
    \item \textbf{Timing and duration of the merger}: Using the H3 main-sequence turn-off sample with precise ages ($\approx10\%$), \citet{Bonaca20} report the star-formation history (SFH) of the ``accreted halo", which is essentially comprised of GSE. The GSE SFH abruptly declines at $\approx10$ Gyr ($z\sim2$). Interestingly, the youngest stars kicked into the in-situ halo are $\approx8$ Gyr old ($z\sim1$). One possible interpretation of these findings is that GSE began interacting with the MW at $z\approx2$ and that the merger concluded by $z\approx1$.

\end{enumerate}
 
\section{Numerical Simulations}
\label{sec:sims}

We aim to reconstruct the GSE merger through controlled, collisionless, N-body simulations. Our strategy is to systematically explore a large grid of simulations spanning reasonable orbital and structural parameters to identify configurations that satisfy the constraints in \S\ref{sec:constraints}. We generate galaxy models for GSE and the MW with the \texttt{GalICv1.1} \citep{Yurin14} initial condition generator and then run merger simulations with the smoothed particle hydrodynamics codes \texttt{Gadget-2} \citep[][]{Springel05} for the initial low-resolution simulations and \texttt{Gadget-4} \citep{Springel20} for the final high-resolution simulations. In all our numerical choices we closely follow recent, similar high resolution merger simulations \citep{Amorisco17,Laporte18,Garavito-Camargo19}. In what follows we motivate the grid we explore and our simulation setup.

\subsection{Structural Parameters}

\label{sec:strucpar}
\subsubsection{GSE}
\label{sec:strucparge}

Our starting point is the GSE stellar mass ($M_{\star} = 2-7 \times 10^{8} M_{\odot}$) and accretion redshift ($z\approx2$) discussed in \S\ref{sec:constraints}. We consider three different models that bracket the literature mass range -- ``M0", ``M1", and ``M2" with stellar masses of $2 \times 10^{8} M_{\odot}$, $5 \times 10^{8} M_{\odot}$, and $7 \times 10^{8} M_{\odot}$ respectively. Extrapolating the size-mass relation (SMR) at $z=2$ from \citet{Mowla19} to lower masses we obtain half-light radii ($r_{\rm{50}}$). At $z\sim2$ half-light radii and half-mass radii are approximately equal \citep[e.g.,][]{Suess19,Mosleh17}. To account for the significant scatter in size at fixed mass observed at $z\sim2$ \citep[e.g.,][]{vanderWel14}, as well as the fact that the SMR has not been measured at masses below $M_{\star}\approx5\times10^{9}$ we consider three sizes: $1\times$, $1.5\times$, and $2\times$ the $r_{\rm{50}}$ from the extrapolated SMR. In total we have nine models for GSE (three stellar masses times three sizes, see Table \ref{table:gestruc}).
 
We model GSE stars as an exponential disk embedded in a spherical, \citet{Hernquist90} DM halo (\texttt{GalIC}'s ``Model D1"). We set the disk scale height to $60\%$ of the disk scale length motivated by simulations that find $z\approx2$ disks are born thick from turbulent gas and stay thick \citep[e.g.,][]{Bournaud09,Forbes12,Bird13,Ma17,Park20}. For the mass of the DM halo we appeal to the $z=2$ stellar mass--halo mass relation from the \texttt{UniverseMachine} empirical model \citep{Behroozi19}. We set the size of the DM halo based on the $z=2$ concentration-mass relation from \citet{Diemer19} that is based on N-body DM simulations. The resulting parameters are listed in Table \ref{table:gestruc}. Note that internally \texttt{GalIC} maps a specified NFW halo mass and concentration to a Hernquist halo of the same mass with a scale length such that the shape of the density profile in the inner regions is identical \citep[][their Eq. 48]{Yurin14}.

\begin{figure*}
\centering
\includegraphics[width=0.9\linewidth]{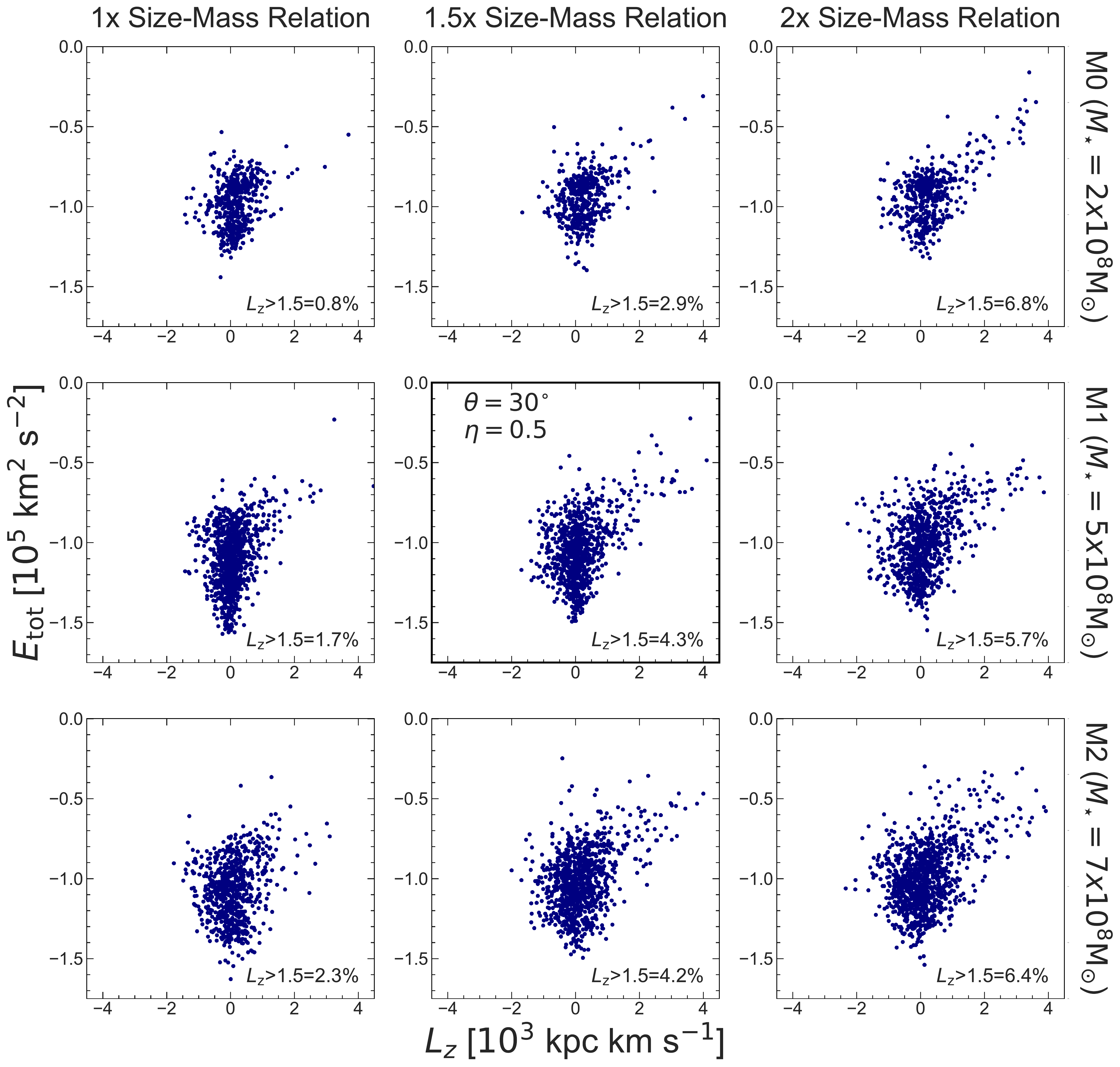}
\caption{Variations in $E-L_{\rm{z}}$ of accreted satellites with stellar mass (rows) and size (columns) while keeping orbital inclination ($\theta=30^{\circ}$) and circularity ($\eta=0.5$) fixed. H3 uncertainties and the survey footprint are applied to the simulation, enabling direct comparison between the two. The central panel shows the most promising model from our initial grid, which reproduces the H3 $r_{\rm{gal}}$ and $L_{\rm{z}}$ distributions. The fraction of debris at $L_{\rm{z}}>1.5$, corresponding to the location of Arjuna, and measured to be $\approx5\%$ in the data, is indicated at the bottom right of each panel. The M0 galaxies (top row) are spread out over a relatively smaller area and do not deposit debris as deep in the potential as the M1 and M2 galaxies (bottom rows), which due to their higher mass rapidly lose energy to dynamical friction. At fixed mass, more extended galaxies produce higher fractions of retrograde debris (compare first and third columns) that arises from their outer, loosely bound regions in the early stages of the merger.}
\label{fig:size-mass}
\end{figure*}

\begin{figure*}
\centering
\includegraphics[width=0.9\linewidth]{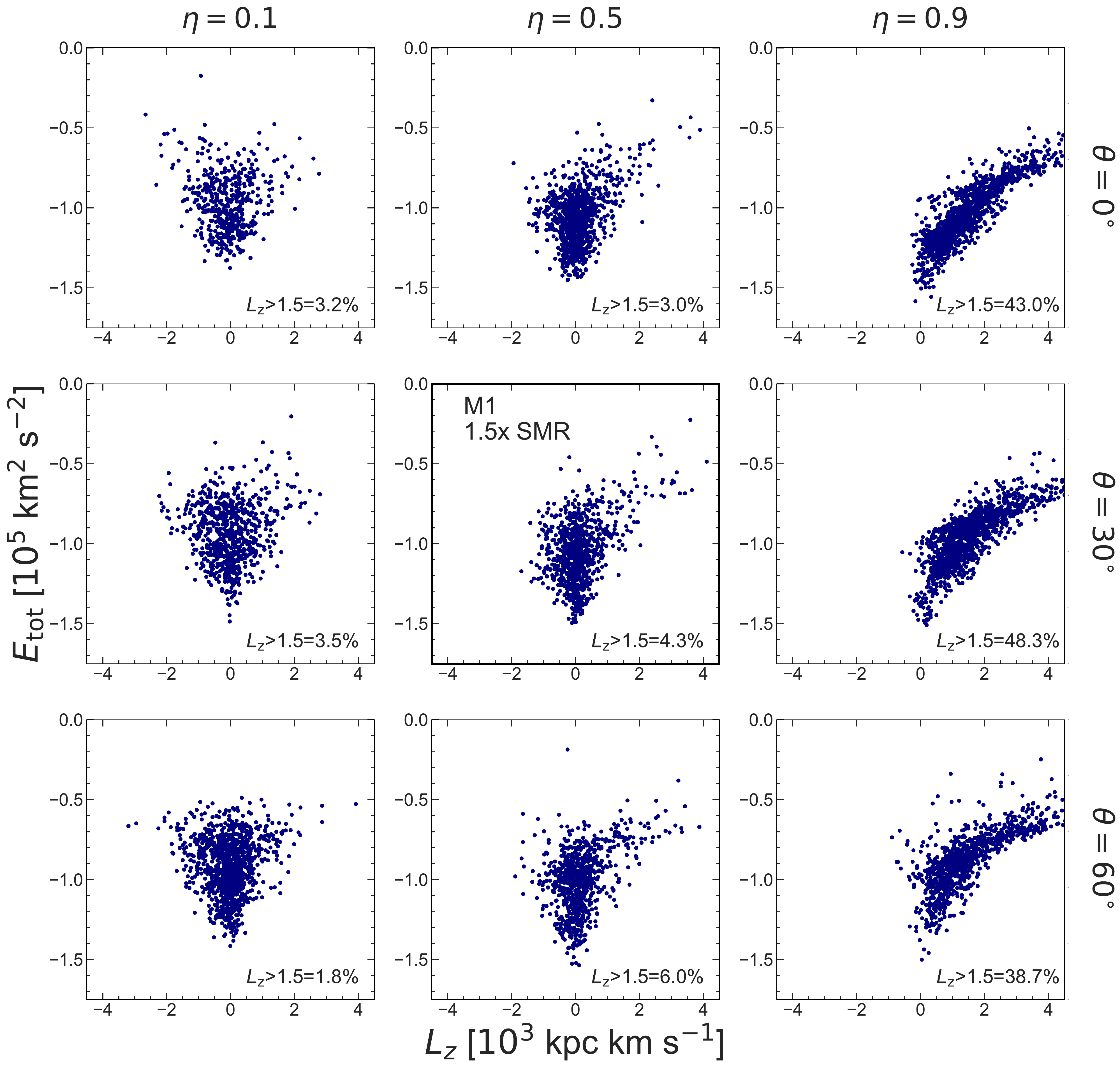}
\caption{Same as Figure \ref{fig:size-mass}, but varying orbital parameters -- the circularity ($\eta$) and inclination ($\theta$) -- while keeping mass and size fixed (M1, 1.5$\times$ SMR). The circularity, $\eta$, strongly influences the retrograde fraction of the observed debris. Circular orbits have longer dynamical friction timescales and so a smaller fraction of their debris is on radial orbits. $\theta$ sets the amount and energy of the debris that would rise into the field of view of a high-galactic field survey like H3. For instance, all the models in the left column produce mergers with indistinguishable orbital decay profiles but $\theta$ sets the number of stars that makes it into the H3 Survey fields.}
\label{fig:eta-theta}
\end{figure*}

\begin{figure*}
\centering
\includegraphics[width=0.9\linewidth]{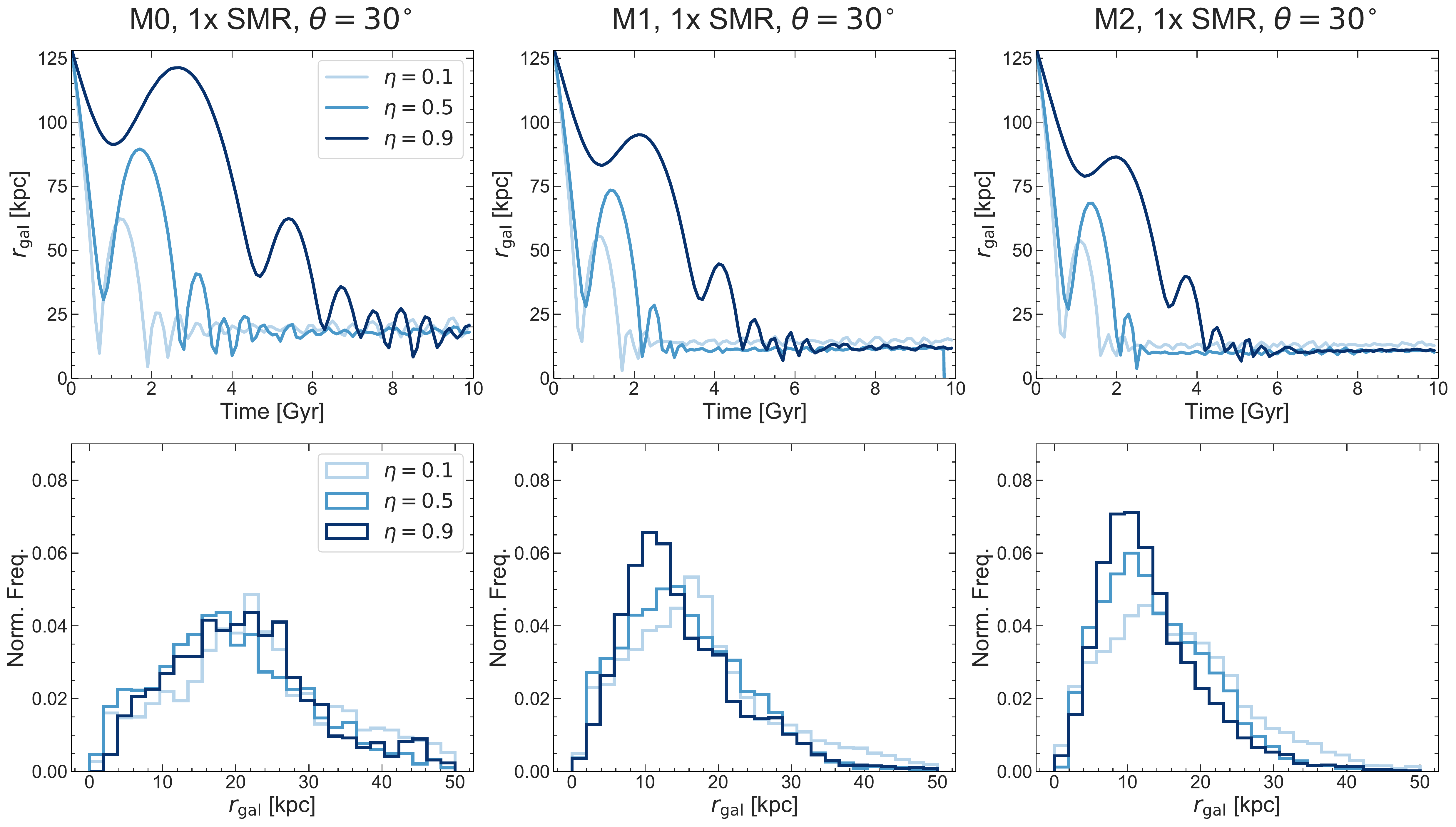}
\caption{\textbf{Top:} Orbital decay as a function of time for mergers of varying mass and circularity. The $5\%$ most bound particles at $t=0$ Gyr are tagged and their center of mass is used to track orbital decay. Lower mass satellites and higher circularity orbits make for prolonged mergers as expected from dynamical friction considerations. \textbf{Bottom:} Spatial distribution of merger debris (all-sky, not limited to H3 footprint). High-circularity, low-mass satellites deposit their stars at larger distances whereas more massive satellites deposit their stars in the inner regions of the host as seen in the distributions growing peakier and shifting left across the panels.}
\label{fig:timing}
\end{figure*}

\begin{figure*}
\centering
\includegraphics[width=0.9\linewidth]{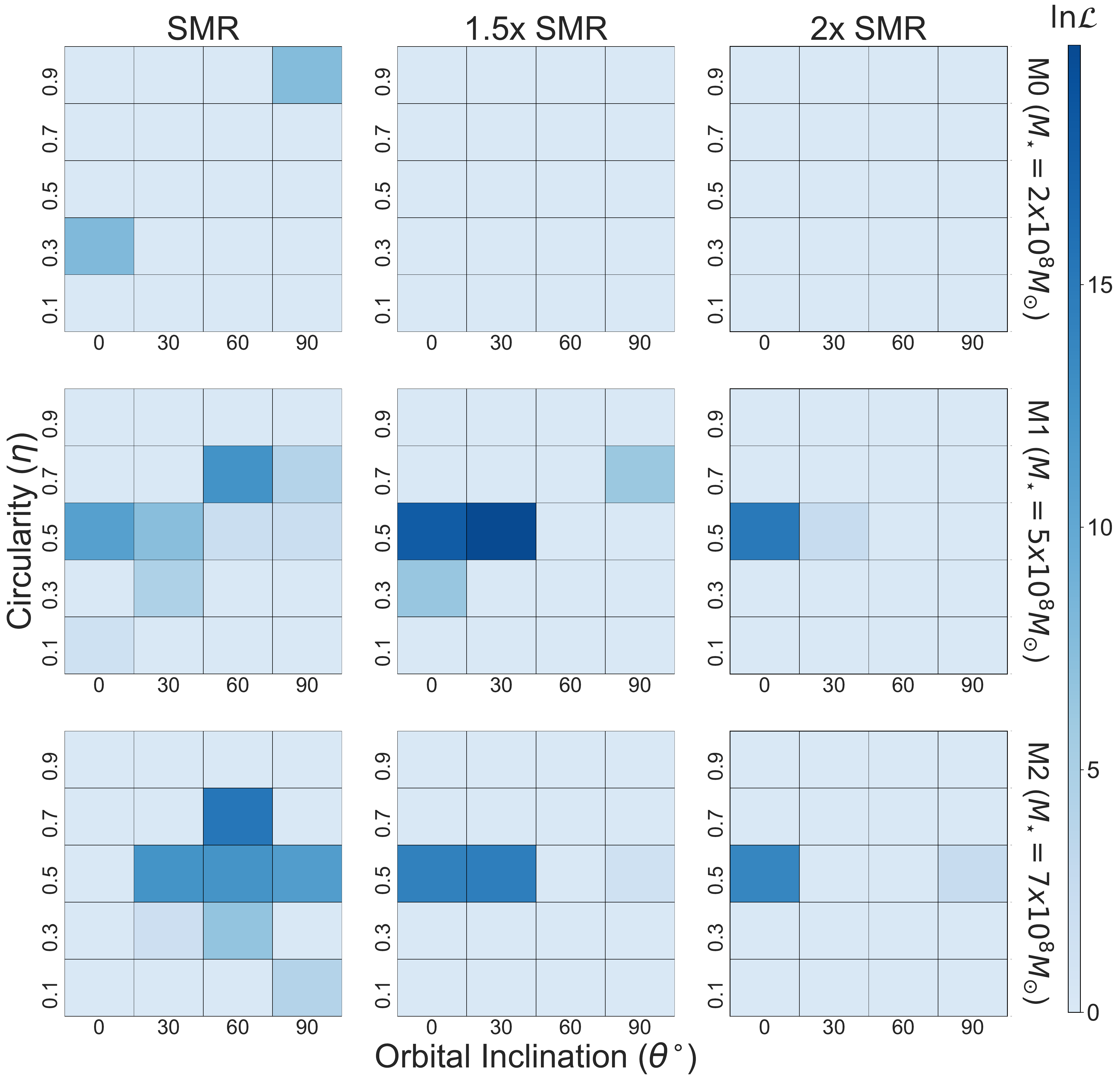}
\caption{Log-likelihood of the initial, low-resolution ($10^{5} M_{\odot}$) grid of retrograde simulations as a function of GSE mass (rows) and GSE size (columns). Each $4\times5$ grid charts circularity ($\eta$) against the orbital inclination ($\theta$). The log-likelihood is computed against the observed H3 $r_{\rm{gal}}$ and $L_{\rm{z}}$ distributions. Remarkably, only these two constraints eliminate the vast majority of the grid. The lowest mass models (M0, top row), regardless of orbital parameters, are heavily disfavored (note the colorbar shows \textit{log}-likelihood). The highest likelihood region occurs in the M1, $1.5\times\rm{SMR}$ simulation at intermediate circularity ($\eta=0.5$) and moderate inclination ($\theta=30^{\circ}$). We re-simulate a finer grid around these parameters at higher resolution ($10^{4} M_{\rm{\odot}}$) and select our fiducial model ($\theta=15^{\circ}$, $\eta=0.5$) from these simulations.}
\label{fig:likelihood}
\end{figure*}

\begin{deluxetable}{lrrrr}
\label{table:gestruc}
\tabletypesize{\footnotesize}
\tablecaption{$z\sim2$ GSE Structural Parameters}
\tablehead{
\colhead{Component} & \colhead{Parameter} & \colhead{M0} & \colhead{M1} & \colhead{M2}}
\startdata
\vspace{-0.2cm}  \\
DM halo & mass, $M_{\rm{200}}$ [$10^{11} M_{\odot}$] & $1.3$ & $2.0$ & $2.5$\\
(Hernquist) & concentration ($c_{\rm{200}}$) & 4.2 & 4.0 & 4.0\\
\hline
Disk & mass [$10^{8} M_{\rm{\odot}}$]  & $2.0$ & $5.0$ & $7.0$\\
(Exponential) & $r_{\rm{50}}$ [kpc] & 
1.5 & 
1.7 & 
1.7 \\
 &  & 
2.3 & 
2.5 & 
2.6 \\
 &  & 
3.0 & 
3.3 & 
3.5 \\
\enddata
\tablecomments{For a model of a given $M_{\rm{*}}$ we consider three scale lengths such that the half-light/half-mass radius, $r_{50}$, is $1\times$ SMR, $1.5{\times}$ SMR, and $2{\times}$ SMR, where SMR is set by the $z=2$ size-mass relation (SMR) in \citet{Mowla19}.}
\end{deluxetable}

\begin{deluxetable}{lrr}
\label{table:mwstruc}
\tabletypesize{\footnotesize}
\tablecaption{$z\sim2$ Milky Way Structural Parameters}
\tablehead{
\colhead{Component} & \colhead{Parameter} & \colhead{Value}}
\startdata
\vspace{-0.2cm}  \\
DM halo & mass, $M_{\rm{200}}$ [$10^{11} M_{\rm{\odot}}$] & $5.0$\\
(Hernquist) & concentration ($c_{\rm{200}}$) & 3.8\\
\hline
Disk & mass [$10^{9} M_{\rm{\odot}}$]  & $6.0$\\
(Exponential) & scale length [kpc] & 
2.0\\
& scale height [kpc] & 1.0\\
\hline
Bulge & mass [$10^{10} M_{\rm{\odot}}$] & $1.4$\\
(Hernquist) & scale length [kpc] & 1.5\\
\enddata
\end{deluxetable}

\begin{deluxetable}{lr}
\label{table:orbitalparams}
\tabletypesize{\footnotesize}
\tablecaption{Orbital parameters explored in simulations}
\tablehead{
\colhead{Property} & \colhead{Parameters}}
\startdata
\vspace{-0.2cm}  \\
\multicolumn{2}{c}{Initial Grid ($m_{\rm{particle}}=10^{5} M_{\rm{\odot}}$)}\\
\hline
Circularity ($\eta$) & 0.1, 0.3, 0.5, 0.7, 0.9\\
Inclination ($\theta$) & $0^{\circ}$, $30^{\circ}$, $60^{\circ}$, $90^{\circ}$\\
Sense of orbit & prograde, retrograde\\
Disk spin & prograde, retrograde\\
\hline
\multicolumn{2}{c}{Refined Grid ($m_{\rm{particle}}=10^{4} M_{\rm{\odot}}$)}\\
\hline
Circularity ($\eta$) & 0.40, 0.45, 0.50, 0.55, 0.60 \\
Inclination ($\theta$) & $0^{\circ}$, $15^{\circ}$, $30^{\circ}$, $45^{\circ}$\\
Sense of orbit & retrograde\\
Disk spin & retrograde\\
\enddata
\end{deluxetable}

\subsubsection{Milky Way}
We require a faithful representation of the MW during the epoch of the merger ($z\approx1-2$). Ages gleaned from a variety of methods suggest that almost the entirety of the present day high-$\alpha$ disk/thick disk as well as the present-day bulge assembled at $z>1$ whereas the low-$\alpha$/thin disk largely grew at $z<1$ \citep[e.g.,][]{Bonaca20,Surot19,Gallart19,Lian20,Ruiz-Lara20}. We therefore model the $z\sim1-2$ MW as a combination of the present-day thick disk and bulge with a total stellar mass of $2\times10^{10} M_{\rm{\odot}}$ and scale lengths following \citet[][]{Bland-Hawthorn16}. Our adopted disk (i.e., the present-day thick disk) has a $\approx30\%$ smaller scale length than the present-day thin disk, accounting for the smaller size of the MW at $z\sim1-2$.  The total stellar mass is consistent with the $z\sim1-2$ expectation from look-back studies of MW progenitors \citep[e.g.,][]{vanDokkum13}.

The disk and bulge are embedded in a spherical, \citet[][]{Hernquist90} DM halo (\texttt{GalIC}'s Model M1). The mass of the DM halo is set to half the $z=0$ mass ($5\times10^{11} M_{\odot}$, \citealt[][]{Vasiliev20, Zaritsky20,Cautun20,Deason21}) motivated by the average growth history of MW-like DM halos seen in simulations \citep[e.g.,][]{Wechsler01}. The concentration is determined by the $z=2$ concentration-mass relation from \citet{Diemer19}.

\subsection{Orbital parameters}
All our simulations begin with GSE at the MW's virial radius. The initial velocity is set such that the total energy is the energy of a circular orbit with radius equal to the MW's virial radius, consistent with satellites in cosmological DM simulations \citep{Jiang15,Amorisco17}. The radial component of the velocity is varied so that the circularity, $\eta$, ranges between 0.1-0.9 in uniform steps of 0.1. Pure radial orbits have $\eta=0$ while perfectly circular orbits have $\eta=1$. The orbital inclination with respect to the MW disk plane, $\theta$, is set to one of $0^{\circ}, 30^{\circ}, 60^{\circ}, 90^{\circ}$. We consider prograde and retrograde orbits and allow the spin of GSE's disk to be co-rotating or counter-rotating with respect to the MW.

\subsection{Merger simulations}
We follow a two-step procedure. We first run $\approx500$ simulations exploring the grid of orbital and structural parameters summarized in Tables \ref{table:gestruc}, \ref{table:mwstruc}, \ref{table:orbitalparams} at a particle resolution of $m_{\rm{DM}} =m_{\rm{baryon}}=10^{5} M_{\odot}$ (``low-res") with the \texttt{Gadget-2} code. We then identify the most promising configurations and simulate another grid around them at a resolution of $m_{\rm{DM}} =m_{\rm{baryon}}=10^{4} M_{\odot}$ (``high-res") with \texttt{Gadget-4} that was released during the course of this project. In the high-res simulations the stellar component of GSE is represented by $\approx$50,000 particles. 

All simulations are run for 10 Gyrs, i.e., from $z=2$ to $z=0$. Time steps ($\Delta t$) are assigned in an adaptive scheme to individual particles via $\Delta t= \sqrt{2\zeta\epsilon/a}$ where $\zeta=0.025$ is an accuracy parameter, $\epsilon$ is the softening length, and $a$ is the gravitational acceleration of the particle under consideration. The softening lengths adopted for all particles are $\epsilon=$250 (80) pc for the low-res (high-res) simulations following the \citet[][]{Power03} criteria for the optimal softening length (their Eqn. 15). The maximum time step is limited to 20 Myr. We choose an opening angle of  $\theta= 0.5^{\circ}$ for the tree algorithm.

\subsection{Comparing models with data}

Since the MW in the simulations is less massive than the present-day MW, we need to account for the deeper $z=0$ potential before comparing with $z=0$ data. Following \citet[][]{Villalobos08} and \citet[][]{Koppelman20b} we measure the mean rotational velocity of the MW disk in our simulations at $2.4\times$ the scale length and compare this with the observed rotation velocity of the MW thick disk at the corresponding distance ($\approx170$ km s$^{-1}$). Based on this comparison we scale our $z=0$ velocities by $1.37\times$ (similar to \citealt[][]{Koppelman20b} who scale by $\approx1.3\times$). This scaling implies there are other sources of mass growth in the inner Galaxy that are not accounted for in our simulation (e.g., the gas from GSE, subsequent accretion events like Sgr, the emergence of the low-$\alpha$ disk). The satisfactory reconstruction of the shape and extent of the $V_{\rm{r}}-V_{\rm{\phi}}$ ``sausage" \citep[][]{Belokurov18} in our fiducial simulation is a consistency check of the applied scaling (see Fig. \ref{fig:fiducial_model} below).

We add observational errors to match the properties of the H3 sample. The PM and RV errors of the sample under consideration are a negligible contribution to the error budget. The distance error is the primary source of uncertainty (see Appendix A of \citealt[][]{Naidu20}). We assume a $10\%$ distance error for all stars, well-matched to the data at hand ($8\pm4\%$). In every simulation snapshot, Galactocentric positions and velocities of stars are computed with respect to the center of mass of the MW bulge stars. Dynamical quantities like angular momenta, energies and Galactocentric velocities are computed exactly as is done for the data in \citet{Naidu20}. For all model-data comparisons, unless otherwise mentioned, we select only the simulation particles with $d_{\rm{helio}}=3-50$ kpc that fall within the survey's fields at $|b|>40^{\circ}$, Dec.$>-20^{\circ}$. Note that the data compared to below are already corrected for the survey selection function (i.e., the photometric magnitude limit and targeting strategy). This means the distribution of say $r_{\rm{gal}}$ or $L_{\rm{z}}$ from the simulations can now be directly compared with the data.

\subsection{General trends}

Here we describe how the various structural and orbital parameters explored in our simulations produce varied debris distributions. We were able to rule out a few regions of parameter space quickly from our initial set of experiments with the M1, SMR model. As one might intuitively expect, none of the prograde simulations produce anything like the strongly retrograde Arjuna debris.  We also found retrograde mergers with counter-rotating disk spin are highly efficient at producing retrograde debris (as seen in e.g., \citealt[][]{Bignone19}). For the rest of this work we focus on such retrograde mergers. Trends with size, mass, and orbital parameters are shown in Figures \ref{fig:size-mass}, \ref{fig:eta-theta}, \ref{fig:timing} to give readers a sense of how these parameters translate into $E-L_{\rm{z}}$ and $r_{\rm gal}$ distributions.

\textit{Size:} At fixed mass, a larger size leads to more retrograde debris (Figure \ref{fig:size-mass}). Since a higher fraction of stars inhabit the outer, less-bound regions of the satellite's disk, they are stripped easily early in the merger. This debris from the outer regions retains memory of the initial (retrograde) orbit of the satellite.

\textit{Mass:} At higher mass, dynamical friction operates more efficiently, satellites sink faster, and deposit a larger fraction of their stars deep in the potential (Figures \ref{fig:size-mass}, \ref{fig:timing}). In particular, $t_{\rm{DF}}\propto M_{\rm{sat}}^{-1}$, where $t_{\rm{DF}}$ is the dynamical friction timescale and $M_{\rm{sat}}$ is the satellite mass \citep[][]{MvdBW}. While quickly radialized, the massive satellites are nonetheless also able to produce a significant fraction of retrograde debris owing to their extended size. Low-mass satellites, on the other hand, experience prolonged mergers and deposit large fractions of their stars at distant radii and higher average energy (top row of Figure \ref{fig:size-mass}).

\textit{Circularity:} Along with mass, $\eta$ is the key moderator of the timing of the merger (Figure \ref{fig:timing}). This is a well-known result:  $t_{\rm{DF}}\propto\eta^{s},\ s\approx0.3-0.5$ \citep[][]{MvdBW}. That is, circular orbits decay the slowest and leave a larger fraction of debris at higher energy/higher angular momenta.

\textit{Inclination}: Once we limit ourselves to retrograde mergers at fixed mass and circularity, the orbital inclination has minimal impact on physical aspects of the merger (e.g., it has little effect on the orbital decay profile). However, $\theta$ strongly moderates the final spatial distribution of the merger debris, and thus how the debris is observed by various surveys (Figure \ref{fig:eta-theta}). For instance, a highly radial, entirely in-plane ($\theta=0$) merger would be barely observable at $|b|>40^{\circ}$ in a survey like H3 but for mergers on inclined orbits the observable fraction is boosted (e.g., by $>3\times$ between $\theta=0^{\circ}$ and $\theta=90^{\circ}$ for a radial $\eta=0.1$ merger).

\begin{figure*}
\centering
\includegraphics[width=0.945\linewidth]{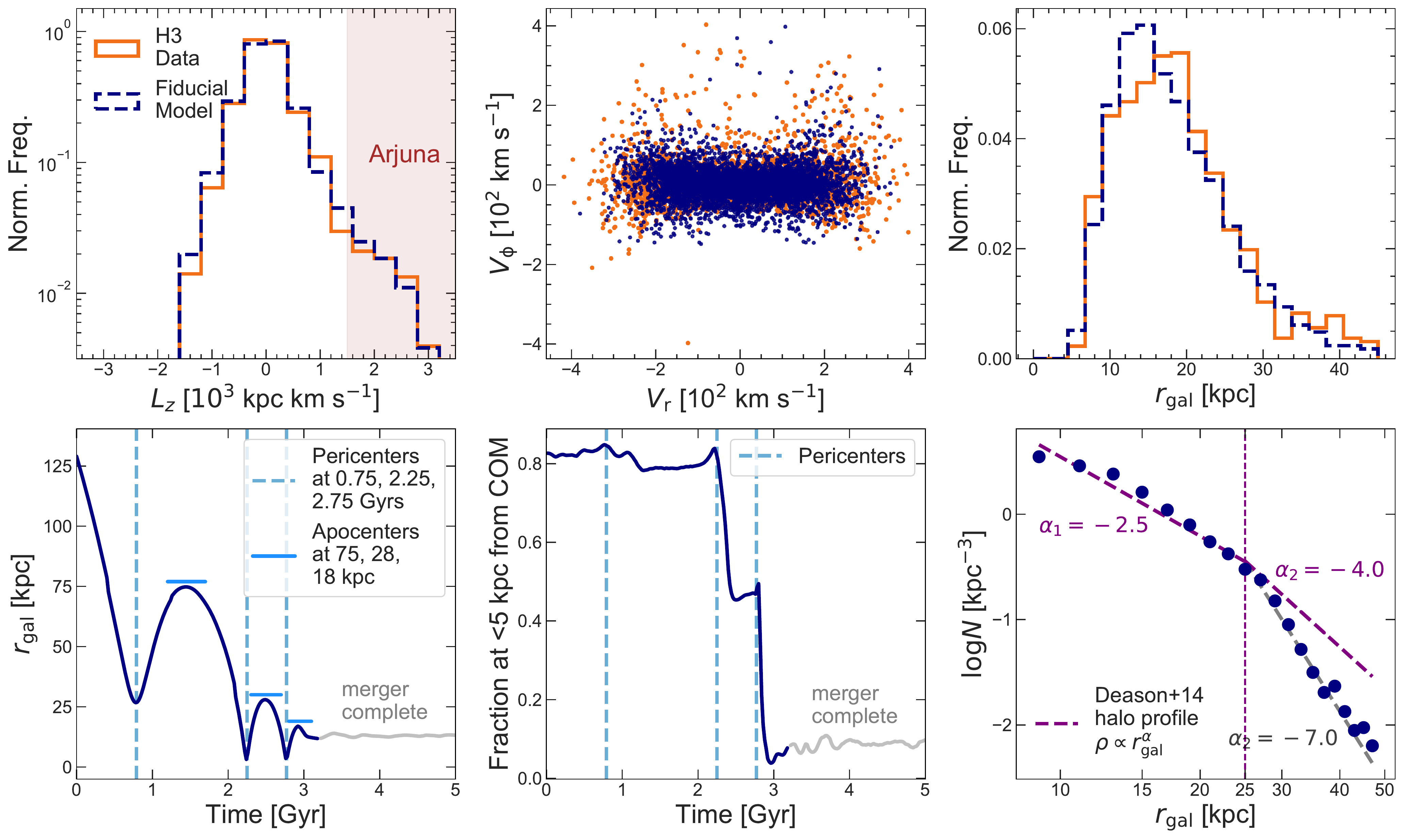}
\caption{Summary of fiducial model. By construction the model is an excellent match to the observed $L_{\rm{z}}$ (\textbf{top left}) and $r_{\rm{gal}}$ distributions (\textbf{top right}). The characteristic $V_{\rm{r}}-V_{\rm{\phi}}$ ``sausage" is satisfactorily reproduced (\textbf{top center}). The orbital decay profile (\textbf{bottom left}) shows a rapid merger. Only 2 Gyrs separate the first and final pericenter. Pericenters (apocenters) are marked by vertical (horizontal) blue lines. In the \textbf{bottom-center} panel we show the fraction of stars within 5 kpc of the COM of the $5\%$ most bound GSE stars with time. The second pericenter is when half the stars are stripped, and the remaining are lost at third pericenter. The all-sky density profile of GSE  (\textbf{bottom-right}) in our model (blue points) and the observed stellar halo profile \citep[][purple]{Deason14} agree well within 25 kpc, and diverge at larger distances where other substructures become important. The break in the profile occurs around the $28$ kpc apocenter in our model.}
\label{fig:fiducial_model}
\end{figure*}

\begin{figure*}
\centering
\includegraphics[width=0.95\linewidth]{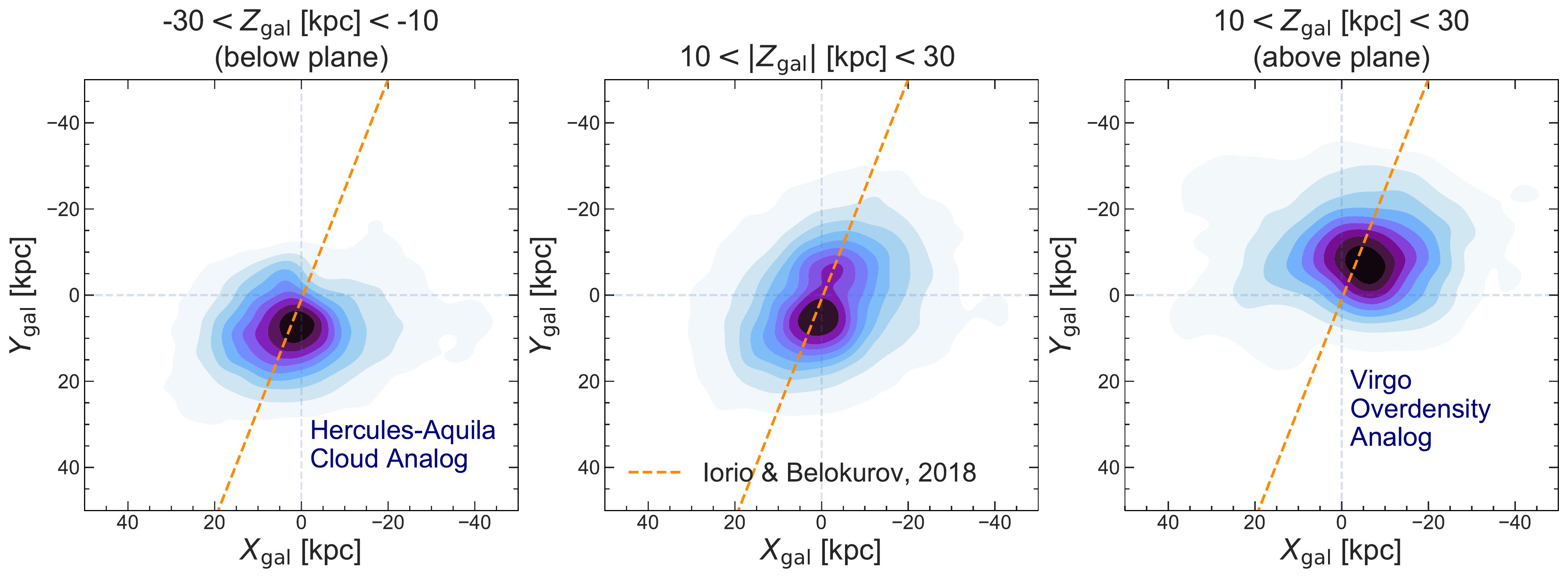}
\caption{Comparison of the fiducial model (blue) with the triaxial inner halo found in \citet[][]{Iorio18,Iorio19} using \textit{Gaia} RR Lyrae. The density contours of our model line up almost exactly with the major axis of their triaxial profile (orange line). Further, the model produces analogs of the Hercules-Aquila Cloud (\textbf{left}) and the Virgo Overdensity (\textbf{right}) at the right locations -- on opposite sides of the plane at either end of the major axis. We emphasize the fiducial model was selected purely on the H3 $L_{\rm{z}}$ and $r_{\rm{gal}}$ and so this close alignment should be viewed as independent corroboration of the model.}
\label{fig:HACVOD}
\end{figure*}

\subsection{Selecting the fiducial model}

To select models that best reproduce the $z=0$ GSE+Arjuna debris we focus on the observed $L_{\rm{z}}$ and $r_{\rm{gal}}$ distributions described in \S\ref{sec:constraints}, and use the other constraints (e.g., the extent of the in-situ halo) as validation checks. $L_{\rm{z}}$ and $r_{\rm{gal}}$ require minimal assumptions (e.g., their computation is independent of the potential), and are sensitive discriminators of merger configurations (Figures \ref{fig:size-mass}, \ref{fig:eta-theta}, \ref{fig:timing}). In detail, we require the $L_{\rm{z}}>1.5$ fraction and $|L_{\rm{z}}|<0.5$ fractions to fall between the 16$^{\rm{th}}$ and 84$^{\rm{th}}$ percentiles of the observed distribution, i.e., the majority of the debris should be radial, but $\approx5\%$ must extend to highly retrograde orbits. We also require the $L_{\rm{z}}<-1.5$ fraction to be $<1\%$ since we observe no stars with GSE chemistry on highly prograde orbits. We make a similar demand of the debris fraction at $r_{\rm{gal}}=10-20$ kpc, $r_{\rm{gal}}=20-30$ kpc, and $r_{\rm{gal}}>30$ kpc. 

The log-likelihood of the entire grid of counter-rotating, retrograde configurations computed against our $r_{\rm{gal}}$ and $L_{\rm{z}}$ requirements listed previously (assuming Gaussian errors) is shown in Figure \ref{fig:likelihood}. In detail, each of the fractions for $L_{\rm{z}}$ (radial, prograde, retrograde) and $r_{\rm{gal}}$ ($10-20$ kpc, $20-30$ kpc, $>30$ kpc) contribute to the likelihood equally as six random normal variables -- this weighting makes for a likelihood that is much more sensitive to the Arjuna component and the break at $25-30$ kpc compared to a classical likelihood computed against the full $r_{\rm{gal}}$ and $L_{\rm{z}}$ distributions that is more sensitive to the peak of the distributions. Interestingly, most of the grid is easily ruled out with our spare set of $r_{\rm{gal}}$ and $L_{\rm{z}}$ constraints. The low mass models (M0) are heavily disfavored no matter their orbital configuration. 

Only one configuration out of the many hundred simulated satisfies these constraints: $M=5\times10^{8} M_{\odot},\ 1.5{\times}\rm{SMR},\  \theta=30^{\circ},\ \eta=0.5$. This configuration has ``Goldilocks" parameters: it is neither too radial nor too circular, has moderate orbital inclination, and an intermediate mass/size. We explore a finer grid around this set of parameters ($\eta=[0.4, 0.45, 0.5, 0.55, 0.6]$, $\theta=[0^{\circ}, 15^{\circ}, 30^{\circ}, 45^{\circ}]$) at $10\times$ resolution (i.e., particle mass of $10^{4} M_{\rm{\odot}}$). We find the $\eta=0.5, \theta=15^{\circ}$ model best matches the data and we focus on this model (the ``fiducial model") for the rest of this work.

We note that there are large swathes of the merger configuration parameter space left unexplored in this work -- e.g., we have kept the MW structural parameters and the initial GSE orbital energy fixed, and we have assumed no scatter in our adopted stellar-mass halo mass relation or mass-concentration relation for GSE. We have also not accounted for the significant amounts of gas GSE likely brought into the Galaxy (potentially $>100\%$ of its stellar mass, \citealt[][]{Tacconi20}) that may have fueled the growth of both the high-$\alpha$ and low-$\alpha$ disks and altered the MW potential \citep[e.g.,][]{Grand20, Bonaca20}. Though note that the overall dynamics of the merger (e.g., the orbital decay profile) are essentially set by the much more massive DM halos of the MW and GSE. As discussed in subsequent sections, our adopted fiducial model is an excellent match to the H3 data and satisfies a wide variety of constraints, but given these caveats we are in no position to claim its parameters are the only ones that match these constraints. 

\section{Results}
\label{sec:results}

\subsection{Preferred configuration: a 2.5:1 merger on an inclined, retrograde orbit}
\label{sec:fiducial}

\begin{figure*}
\centering
\includegraphics[width=0.95\linewidth]{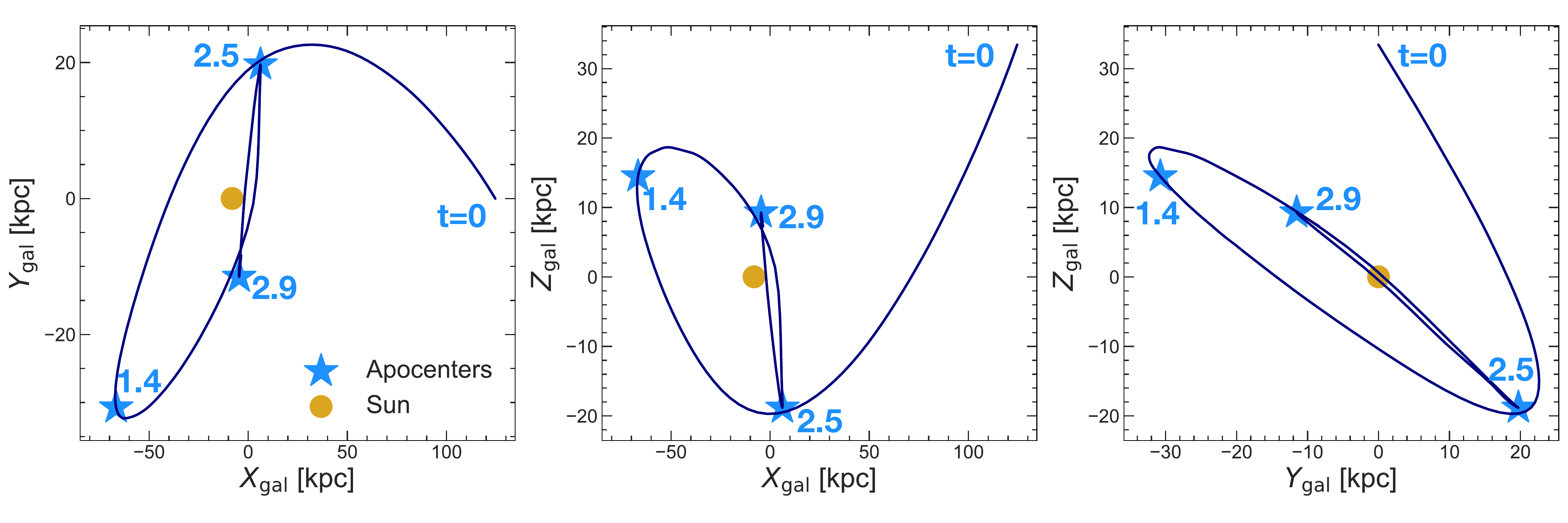}
\caption{Orbit of fiducial model. Apocenters are marked as stars along with their time of occurrence in Gyrs.}
\label{fig:orbits}
\end{figure*}

\begin{figure*}
\centering
\includegraphics[width=\linewidth]{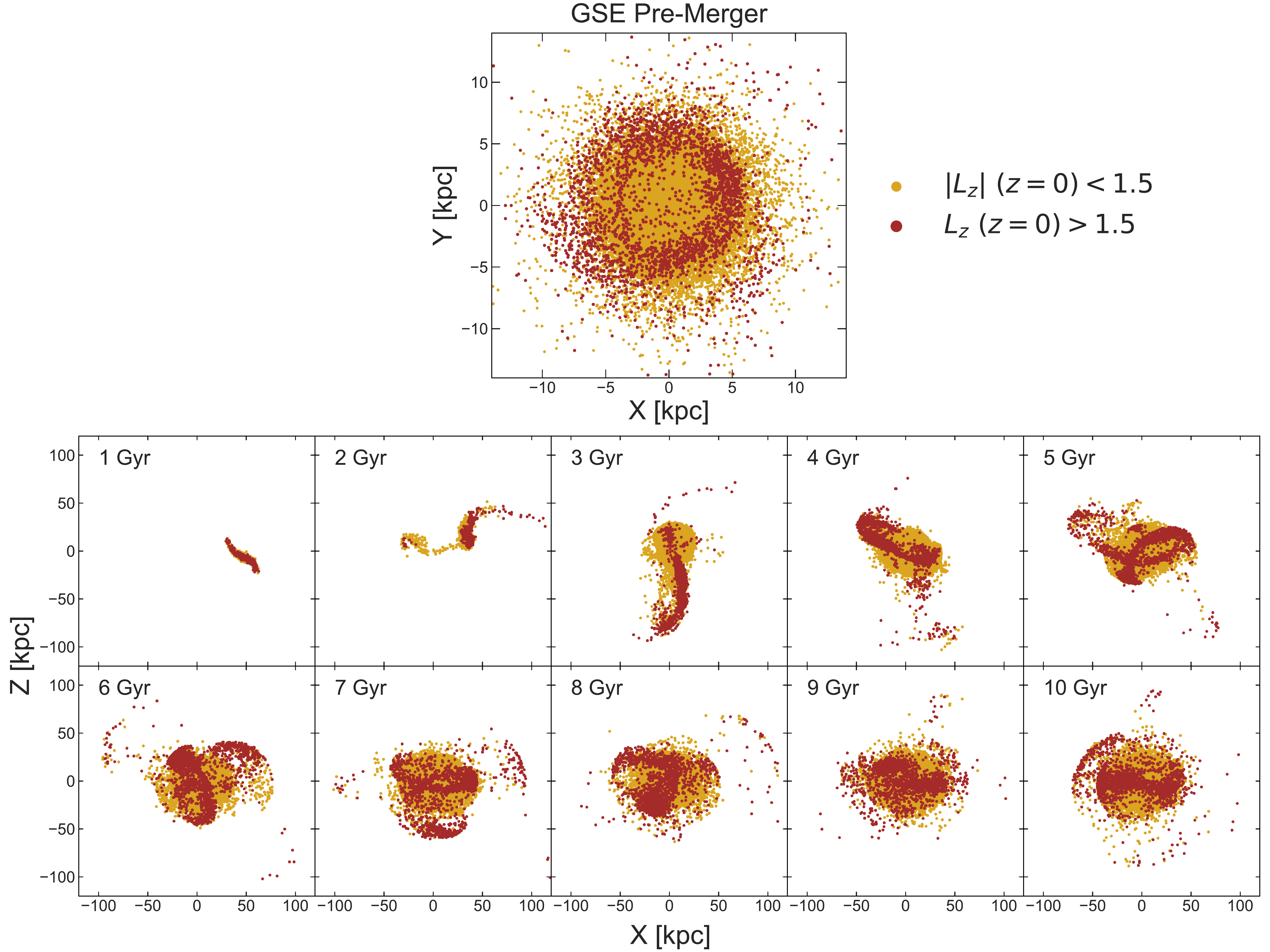}
\caption{The origin of Arjuna according to the fiducial model. \textbf{Top:} The GSE disk before the merger. Stars are colored brown if they end up on highly retrograde, Arjuna-like orbits ($L_{\rm{z}}>1.5$), and gold otherwise. The Arjuna-like stars preferentially inhabit the outer disk (${<}r_{\rm{gal}}{>}\approx2.5\times r_{\rm{50}}$). \textbf{Bottom:} Progression of the merger in the $X$-$Z$ plane. Arjuna stars are stripped early, at larger distances, when the satellite still has its initial retrograde angular momentum. The other stars are stripped after the satellite is radialized, within a $\approx25$ kpc golden ball which corresponds to the second apocenter.}
\label{fig:arjuna_origin}
\end{figure*}

The fiducial merger configuration ($M_{\star}=5\times10^{8} M_{\odot},\  M_{\rm{DM}}=2\times10^{11} M_{\odot},\  1.5{\times}\rm{SMR},\ \theta=15^{\circ},\ \eta=0.5$) selected from the high-resolution grid is summarized in Figure \ref{fig:fiducial_model}. The $L_{\rm{z}}$ and $r_{\rm{gal}}$ distributions are an excellent match to the H3 data by construction. The ``sausage" in $V_{\rm{r}}-V_{\rm{\phi}}$ where GSE was first discovered with \textit{Gaia} is satisfactorily reproduced \citep[][]{Belokurov18}. The orbit has an apocenter at $28$ kpc which is where several studies have found a break in the density profile in the halo. The slope of the GSE density profile within $25$ kpc is a good match to that found for the inner halo \citep[e.g.,][discussed further in \S\ref{sec:profile}]{Deason14}. This is exactly as expected given that GSE dominates the halo within 25 kpc \citep[e.g.,][]{Naidu20}. At larger distances, other components become more prominent, and the GSE density profile falls faster than that of the overall halo. The merger is fairly rapid, with the gap between first and final pericenter being a mere 2 Gyrs, in agreement with the timing constraints (discussed further in \S\ref{sec:timing}). We also confirm that $>90\%$ of the stars kicked out of the Milky Way disk (the in-situ halo) are contained within $|Z|<15$ kpc. 

In Figure \ref{fig:HACVOD} we show that the fiducial model also produces overdensities analogous to the Hercules-Aquila Cloud and Virgo Overdensity at their exact observed locations. Furthermore, the direction along which the debris is spread out is an excellent match to that of the major axis of the triaxial ellipsoid fit by \citet{Iorio19} to describe the inner halo. The agreement between our fiducial model and these spatial constraints is particularly remarkable since we select the model only based on the H3 $L_{\rm{z}}$ and $r_{\rm{gal}}$ distributions (implicitly, the H3 window function has a spatial aspect). Now that we have a model which is in excellent agreement with the constraints in \S\ref{sec:constraints}, we can use it to extract further physical insights about the merger.

The total GSE mass of our fiducial model ($2\times10^{11} M_{\odot}$) is $\approx50\%$ higher than that of the Large Magellanic Cloud ($\approx1.3\times10^{11} M_{\odot}$, \citealt[][]{Erkal19,Vasiliev20}) and represents as much as $20\%$ of the MW's present-day virial mass ($\approx10^{12} M_{\odot}$, \citealt[][]{Zaritsky20,Cautun20,Deason21}). The stellar mass constitutes $\approx50\%$ of the MW's stellar halo ($\approx10^{9} M_{\odot}$, \citealt[][]{Deason19,Mackereth20}). Within the ambit of our grid, this finding is particularly robust, since the lower mass GSE models are strongly ruled out by the $r_{\rm{gal}}$ and $L_{\rm{z}}$ constraints (top row, Figure \ref{fig:likelihood}) -- these models produce slowly decaying mergers that deposit a high fraction of their debris at larger distances than seen in the data (Figure \ref{fig:timing}).

In Figure \ref{fig:orbits} we plot the orbit of GSE from the fiducial model. This orbit is computed based on the centre of mass of the $5\%$ most bound GSE stars prior to the merger and shown for the first 3 Gyrs of the simulation (i.e., covering the duration of the merger). The orbit is not radial right away -- for $\approx$2 Gyrs GSE journeys through the Galaxy with significant angular momentum before ending up on a radial track at $<30$ kpc between its final two apocenters (shown as stars). We will refer to this orbit at various points in subsequent sections while interpreting e.g., the spatial distribution of the GSE debris.

\subsection{The origin of Arjuna}
\label{sec:arjuna_origin}
In Figure \ref{fig:arjuna_origin} we trace the origins of the highly retrograde Arjuna stars. We tag stars that are at $L_{\rm{z}}>1.5$ at $z=0$ and follow them through the simulation. In the top panel of Figure \ref{fig:arjuna_origin} we see that before the merger, these stars occupy the outer regions of the disk of GSE and lie at a median radius of $\approx2.5\times r_{\rm{50}}$. These relatively loosely bound stars from the outer disk are stripped earlier in the merger, and so they retain the larger angular momenta and higher energy that the satellite initially arrived with. On the other hand, the majority of stars from the inner disk are stripped after the bulk motion of the satellite has been radialized and hence they are found on $|L_{\rm{z}}|<0.5$, eccentric orbits and appear as the $V_{\rm{r}}-V_{\rm{\phi}}$ ``sausage". An implication of this exercise is that information about the detailed spatial structure of a galaxy that was disrupted $\approx10$ Gyrs ago is still retained in the present-day angular momenta distribution of its debris in the halo. 

\begin{figure*}
\centering
\includegraphics[width=0.9\linewidth]{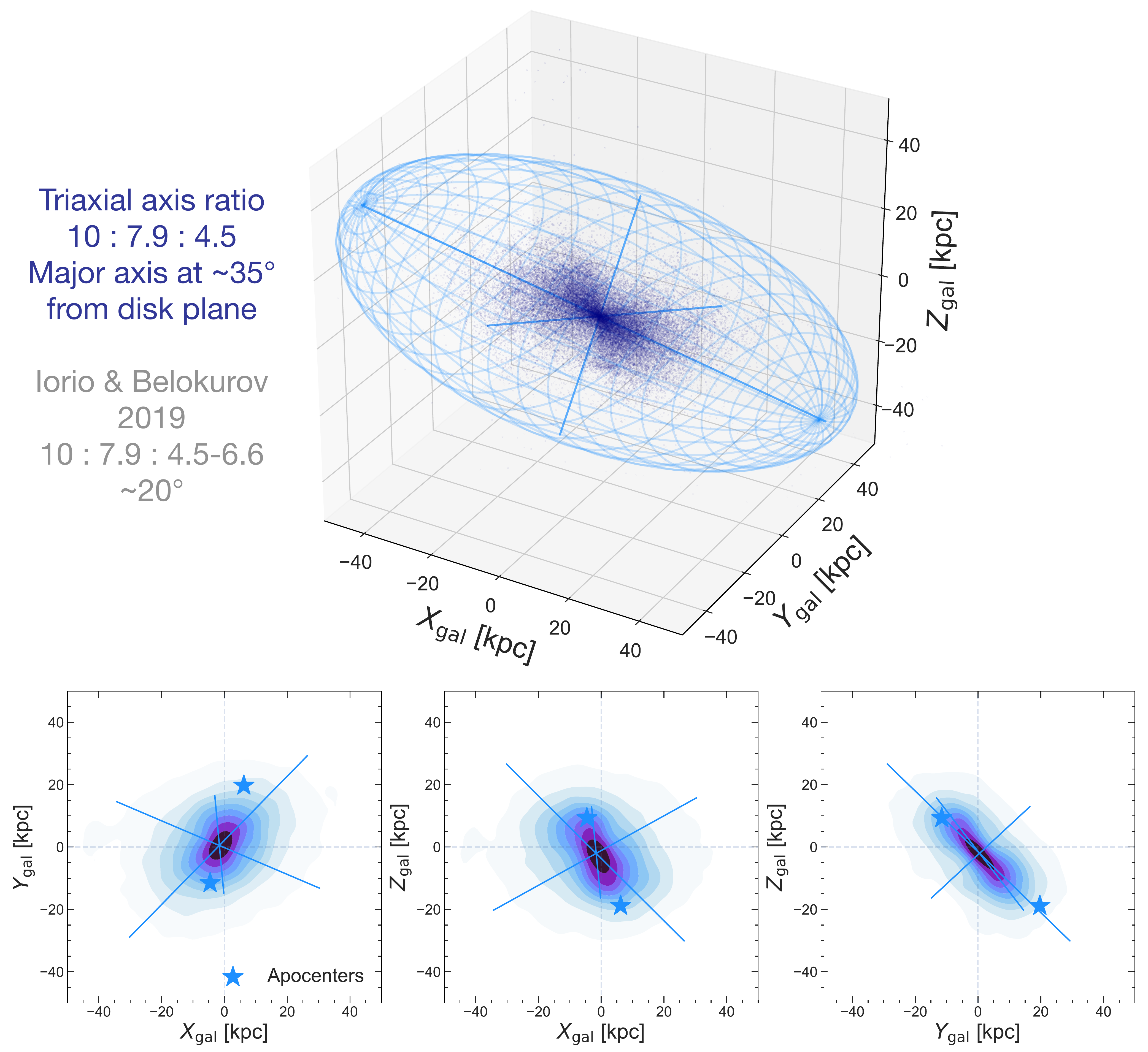}
\caption{Shape of the MW inner halo in the fiducial model. \textbf{Top:} Within 35 kpc, GSE is by far the most dominant component of the halo, and so the geometry of its debris sets the shape of the inner halo. We fit a trixial ellipsoid (light blue grid) to describe the GSE debris (dark blue points). The major axis sticks out of the Galactic plane at $\approx35^{\circ}$. Our derived triaxial halo parameters agree very well with those found using \textit{Gaia} RR Lyrae \citep[][]{Iorio18,Iorio19} even though no shape information is used to constrain the model. \textbf{Bottom:} 2D projections of the stellar density with the ellipsoid axes overplotted. In each panel one of the three axes closely tracks the debris density. The tilt of the ellipsoid out of the plane and its elongated morphology is clearly seen in the bottom-right panel. The major axis in this panel tracks almost exactly the line joining the penultimate (2.5 Gyr) and final (2.9 Gyr) apocenters depicted as stars.} 
\label{fig:shape}
\end{figure*}

\begin{figure*}
\centering
\includegraphics[width=0.9\linewidth]{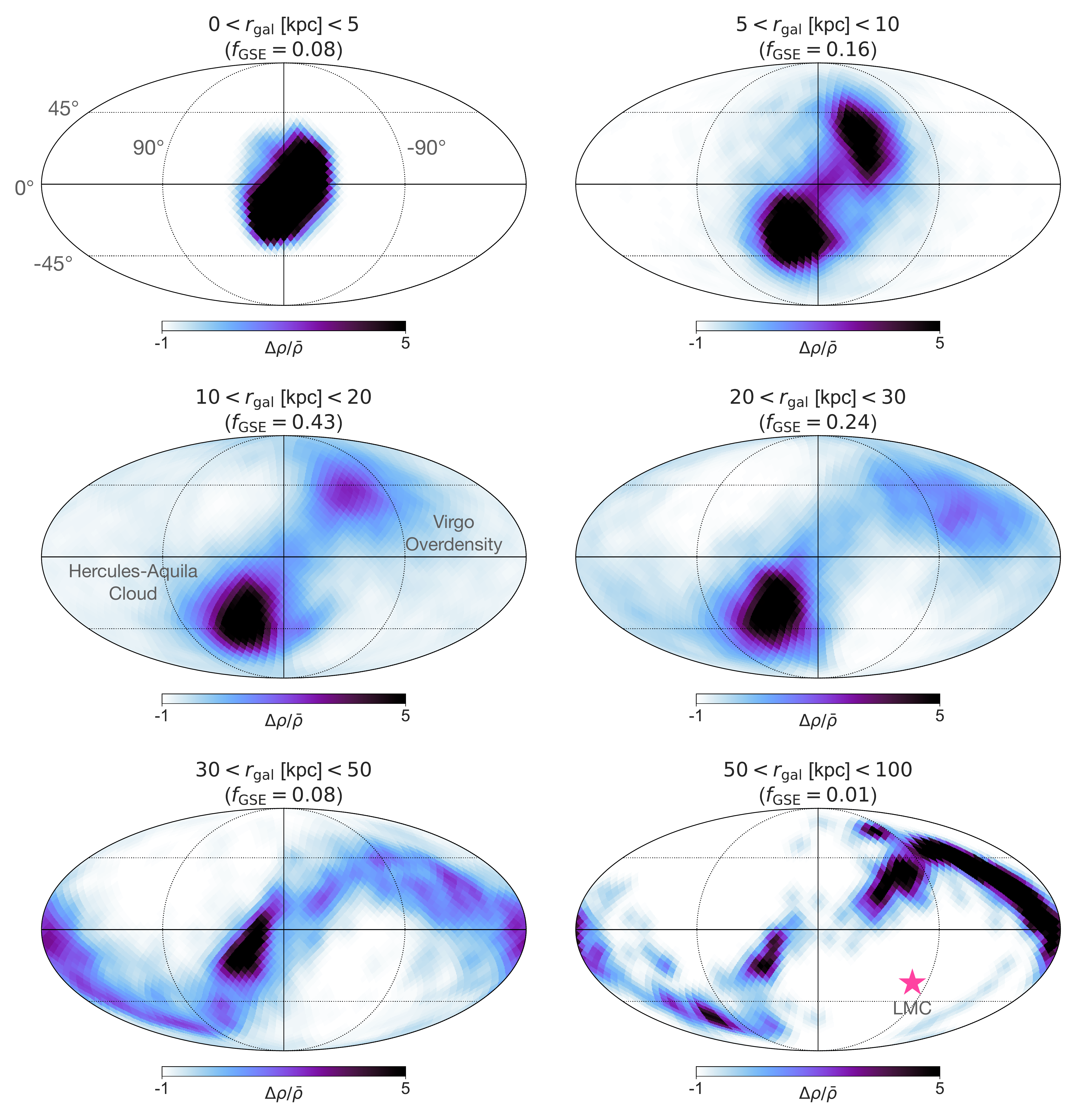}
\caption{All-sky debris density maps from the fiducial model in  Molleweide projection and Galactic coordinates smoothed with an FWHM=$10^{\circ}$ kernel. We show bins in Galactocentric distance ($r_{\rm{gal}}$) between 0-100 kpc with the fraction of debris in each bin ($f_{\rm{GSE}}$) indicated in the title. These maps are richly structured. At $r_{\rm{gal}}<20$ kpc two prominent lobes are apparent, one above the plane, and one below -- these correspond to the locations of the Hercules-Aquila Cloud and the Virgo Overdensity. The northwest and southeast quadrants contain the bulk of GSE  stars at all distances, underscoring the strong spatial anisotropy of the debris. The position of the LMC is indicated with a pink star in the bottom right panel.}
\label{fig:allsky}
\end{figure*}

\begin{figure*}
\centering
\includegraphics[width=0.9\linewidth]{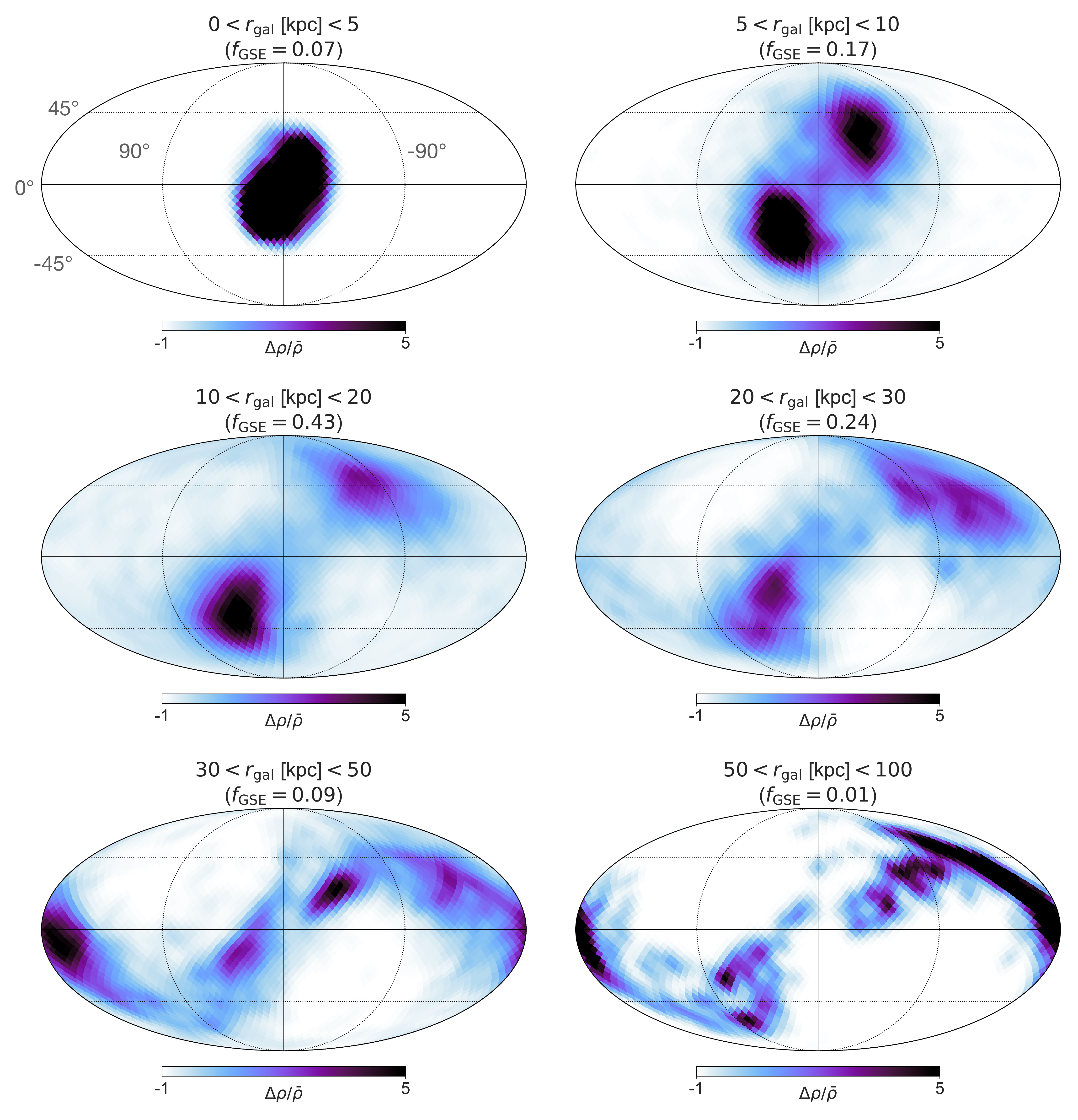}
\caption{Same as Figure \ref{fig:allsky}, but 500 Myrs earlier ($t=9.5$ Gyrs) in the simulation. While the regions of the sky inhabited by GSE debris are largely unchanged, and the integrals of motion (such as $L_{\rm{z}}$) are stable, the relative density in these regions fluctuates as stars orbit between them. These density fluctuations are particularly dramatic in the $20<r_{\rm{gal}}\ \rm{[kpc]}<30$ (center-right) and $30<r_{\rm{gal}}\ \rm{[kpc]}<50$ (lower-left) bins (compare with Fig. \ref{fig:allsky}).}
\label{fig:allsky2}
\end{figure*}

\begin{figure*}
\centering
\includegraphics[width=0.9\linewidth]{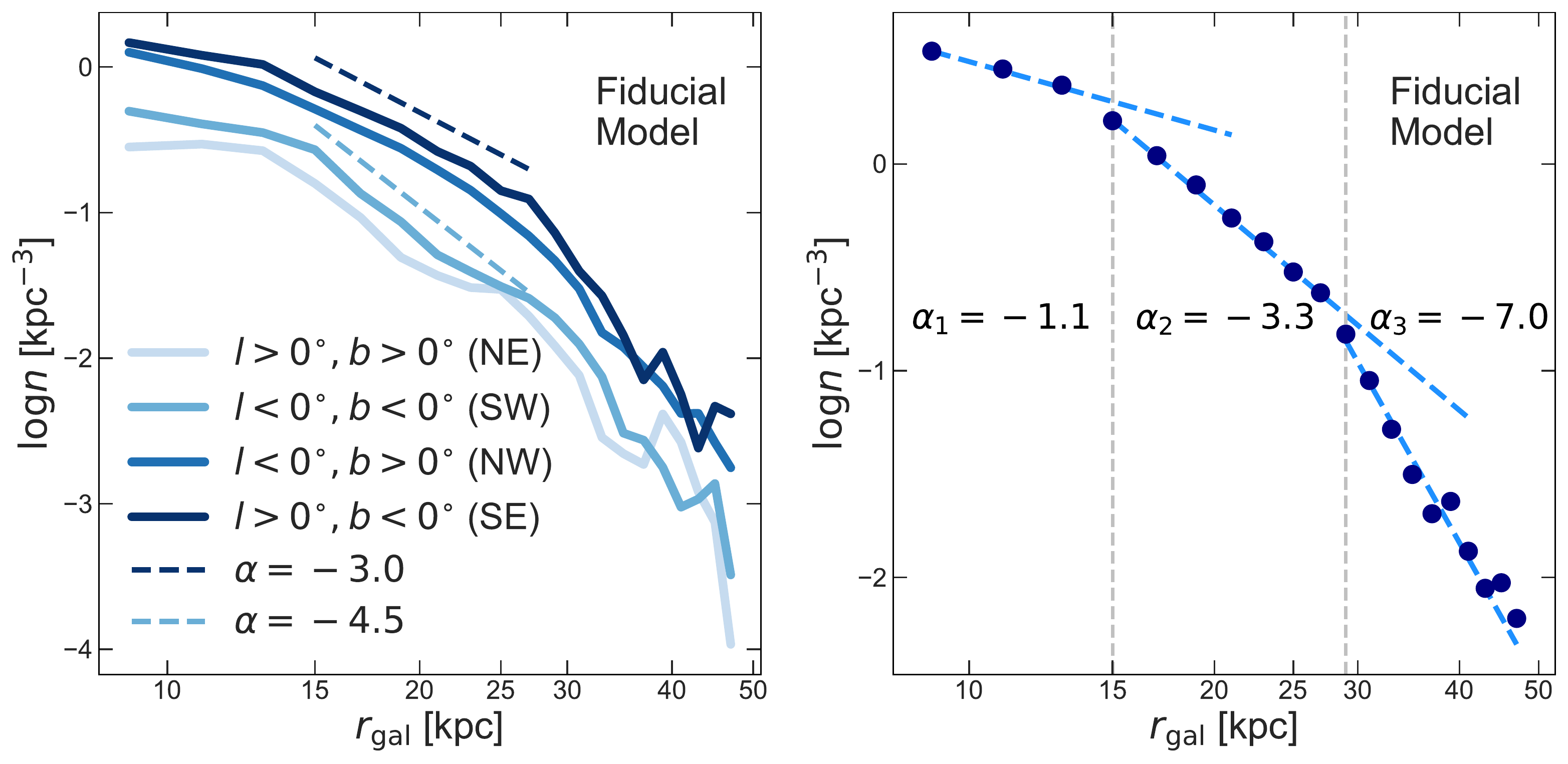}
\caption{\textbf{Left:} Variation of GSE density profile with on-sky location according to the fiducial model. There is a $5-10\times$ lower density in the NE and SW quadrants as a consequence of the triaxial GSE debris distribution (see Fig. \ref{fig:allsky}). Further, the density profile and location of the breaks in the profile shift across the sky -- we highlight the different slopes in the 15-25 kpc range for the southeast ($\alpha=-3.0$) and southwest ($\alpha=-4.5$) quadrants of the sky. These variations have important implications for halo density profile measurements that typically probe only a fraction of the sky. \textbf{Right:} We propose a ``double-break" all-sky density profile for GSE with one break at $\approx15-18$ kpc and another at $\approx30$ kpc. This profile is motivated by the location of the apocenters in our simulation (Figure \ref{fig:fiducial_model}, \ref{fig:orbits}).}
\label{fig:profile}
\end{figure*}

\subsection{The shape of the inner halo}
\label{sec:shape}

A large body of literature has pursued the morphology of the stellar halo since it is expected to be a reasonable tracer of the much more massive dark matter halo \citep[e.g.,][]{Newberg06, Miceli08,Watkins09,Sesar13,Posti19}. With \textit{Gaia} it has become apparent that the inner halo ($<30$ kpc) is essentially built out of GSE and so these studies were in fact measuring the distribution of GSE debris in great detail. Here we connect our fiducial model to the halo morphology literature.

In Figure \ref{fig:shape} we show a triaxial ellipsoid fit to the simulated debris distribution. We first fix the orientation of the orthogonal axes via principal component analysis and then use the \texttt{nestle}\footnote{https://github.com/kbarbary/nestle} ellipsoid bounding routine to measure the relative axes ratios. To ensure the robustness of the fit we remove the most distant $1\%$, $5\%$ and $15\%$ of the debris and find the axes ratios are stable to $<10\%$. The ellipsoid is centered on the Galactic center. The orientation of the axes is described by the rotation matrix $R(\gamma, \beta, \alpha)= R_{\rm{Z}}(\gamma)R_{\rm{Y}}(\beta)R_{\rm{X}}(\alpha)$ where $\gamma\approx-35^{\circ}$, $\beta\approx-5^{\circ}$, $\alpha\approx-135^{\circ}$ are counter-clockwise yaw, pitch, and roll angles respectively. The axes ratios, in terms of the pre-rotation axes are $X:Y:Z = 7.9 :10:4.5$. The major axis of the ellipsoid is at $\approx35^{\circ}$ to the plane. Perched on either end of the major axis are overdensities analogous to the Hercules-Aquila Cloud and the Virgo Overdensity. To visualize this debris geometry, imagine the $Y=X$ line and then lift it out of the plane by $\approx35^{\circ}$ such that it points from (+Y, -Z) to (-Y, +Z). 

The elongated morphology of GSE debris is particularly evident in the Y-Z plane where GSE stars almost entirely lie in two quadrants (bottom-right, Figure \ref{fig:shape}). This geometry is set by the locations of the final two apocenters that the bulk of stars are stripped between. The apocenter locations are highlighted as stars in the bottom panel of Figure \ref{fig:shape} -- one lies above the plane, the other below the plane. The major axis of the triaxial ellipsoid closely tracks the line joining these two apocenters.

These derived structural parameters are in good agreement with the triaxial ellipsoid model of \citet{Iorio18,Iorio19}, who inferred the shape of the inner stellar halo ($<30$ kpc) using a homogeneously selected all-sky dataset (\textit{Gaia} RR Lyrae) for the first time. These authors report remarkably similar axes ratios to those measured in our fiducial simulation  ($ 7.9 : 10: 4.5-6.6$; they allow the minor axis ratio to vary with distance). Further, they find the halo is at a $20^{\circ}$ angle to the disk plane.

As hinted in \citet[][]{Iorio19} we note that the major axis of GSE debris points in the same direction that the Magellanic Clouds entered the Galaxy from (based on the LMC orbit in \citealt[][]{Garavito-Camargo19}). A tantalizing possibility is that both GSE and the LMC traveled along the same cosmic web filament that feeds the Milky Way. This hypothesis would be strengthened if GSE merged on a purely radial orbit along the major axis. However, our fiducial model disfavors this scenario: the retrograde GSE enters the MW from a different direction and eventually ends up along the major axis only after being radialized (see Figure \ref{fig:orbits}).

\subsection{GSE throughout the halo}
\label{sec:outerhalo}

Figures \ref{fig:allsky} and \ref{fig:allsky2} present all-sky density maps of GSE debris from our fiducial model, split in radial bins. The defining feature of these maps is that the GSE debris is structured and far from isotropic/axisymmetric. Fig. \ref{fig:allsky} depicts the debris at $z=0$ in our model while Fig. \ref{fig:allsky2} shows the debris 500 Myrs ago. Comparing these figures shows that while the regions occupied by GSE debris are stable, their relative densities fluctuate as stars orbit between these regions. Detailed density comparisons with all-sky data must take this time variability into account. 

In the inner halo ($r_{\rm{gal}}<30$ kpc) the GSE debris is spread across an inclined axis that runs through $l=0^{\circ}, b=0^{\circ}$. This is the major axis of the ellipsoid fit in \S\ref{sec:shape}. Occurring on either end of it are overdensities that correspond to the HAC and VOD. Stars in the present-day HAC comprised the VOD 0.5 Gyrs ago, and vice versa. The HAC/VOD are where the stars slow down and come to a halt before they turn around to descend/ascend the plane and so these are the regions where stars pile up into on-sky overdensities. 

Beyond 30 kpc, GSE stars trace stream-like patterns across the sky (bottom panels of Figs. \ref{fig:allsky}, \ref{fig:allsky2}) and are retrograde ($\langle L_{\rm{z}}\rangle (r_{\rm{gal}}>30\ \rm{kpc})=1.1$, $\langle L_{\rm{z}}\rangle (r_{\rm{gal}}>50\ \rm{kpc})=2.3$). This stream-like debris at $>30$ kpc arises from $2-3\times r_{\rm{50}}$ in the GSE disk. We predict all-sky maps of metal-rich ([Fe/H]$\approx-1.2$), retrograde stars at these distances will show the diffuse ``leading arm" and ``trailing arm" of GSE seen in the bottom two panels. Without velocity information, this detection might be made challenging by the on-sky overlap with Sgr -- $\approx50\%$ of GSE debris beyond 30 kpc is at $|B_{\rm{Sgr}}|<20^{\circ}$, where $B_{\rm{Sgr}}$ is latitude in the Sgr plane defined in \citet[][]{Belokurov14}. The $|B_{\rm{Sgr}}|<20^{\circ}$ fraction rises to $\approx75\%$ when considering the $|l|>90^{\circ}$ regions. However, with proper motions and velocities, distinguishing between the highly retrograde/radial GSE debris and the prograde Sgr debris that has high $L_{\rm{y}}$ \citep[][]{Johnson20} will be trivial.

Also indicated in the bottom-right panel is the location of the LMC. The LMC, due to its significant mass, and because it is on first infall is predicted to induce large-scale features across the sky. In particular, \citet[][]{Garavito-Camargo19} forecast a ``collective response" overdensity in the northern hemisphere, and a dynamical friction wake overdensity in the south-east quadrant at $r_{\rm{gal}}\geq45$ kpc. These quadrants are predicted to also harbor GSE debris ($\approx10\%$ of the total mass) at $r_{\rm{gal}}=30-100$ kpc.  Similarly, efforts to constrain the barycentric motion of the MW due to the LMC by comparing radial velocities in the northern and southern hemispheres could be impacted by GSE stars at these distances \citep[e.g.,][]{Erkal20}. At $r_{\rm{gal}}=40-100$ kpc, the northern GSE stars have $V_{\rm{GSR}}\approx85$ km s$^{-1}$ and the southern stars have $V_{\rm{GSR}}\approx-65$ km s$^{-1}$, i.e., the radial velocities of GSE stars mimic the expected LMC-induced redshift and blueshift signals.  In detail these signals should be separable both because the predicted GSE debris is confined to relatively cold streams on-sky and because the predicted proper motion signals will differ.

\subsection{A second apocenter at $\approx$15 kpc and a ``double-break" inner halo profile}
\label{sec:profile}

In Figure \ref{fig:profile} we examine the density profile of GSE as a function of sky position (left panel) and integrated over the sky (right panel). We propose a ``double-break" profile for the inner halo with a prominent break at the penultimate apocenter ($\approx28-30$ kpc) of the GSE orbit, and another break close to its final apocenter ($\approx15-18$ kpc). Since GSE is by far the most dominant component of the inner halo ($r_{\rm{gal}}<30$ kpc), we expect the overall halo density profile at these distances to largely trace the GSE profile.

Interestingly, several studies have found a ``single-break" profile for the inner halo \citep[e.g.,][]{Watkins09,Deason14,Xue15}. These single-break profiles do provide a reasonable fit to our model -- an example \citep[][]{Deason14} is shown in the bottom-right panel of Figure \ref{fig:fiducial_model}. We also observe that the ``single-break" inner halo profiles in the literature are divided about the location of the break, with some favoring $\approx25-30$ kpc \citep[e.g.,][]{Watkins09,Sesar11,Deason11,Faccioli14} and others finding $\approx15-20$ kpc \citep[e.g.,][]{Sesar13,PilaDiez15,Xue15}. This unsettled state of affairs may be due to the fact that the halo density profile is not being well-represented by a ``single-break" function and also that it shows large scale variation across the sky (left panel of Figure \ref{fig:profile}). The variation is expected from the tilted ellipsoid geometry of the debris -- not only does the normalization of the profile vary by $\approx5-10\times$, but also the shape of the profile shifts significantly from region to region. Our simulation motivates remeasuring the halo density profile allowing for an extra break. We provide power-law coefficients for our proposed $\rho \propto r_{\rm{gal}}^{\rm{\alpha}}$ profile as a promising, physically motivated launching point for future measurements: $\alpha\ ({<}15\ \rm{kpc})=-1.1$, $\alpha\ (15-30\ \rm{kpc})=-3.3$.

\subsection{Interpreting the timeline of the GSE merger}
\label{sec:timing}
\begin{figure}
\centering
\includegraphics[width=\linewidth]{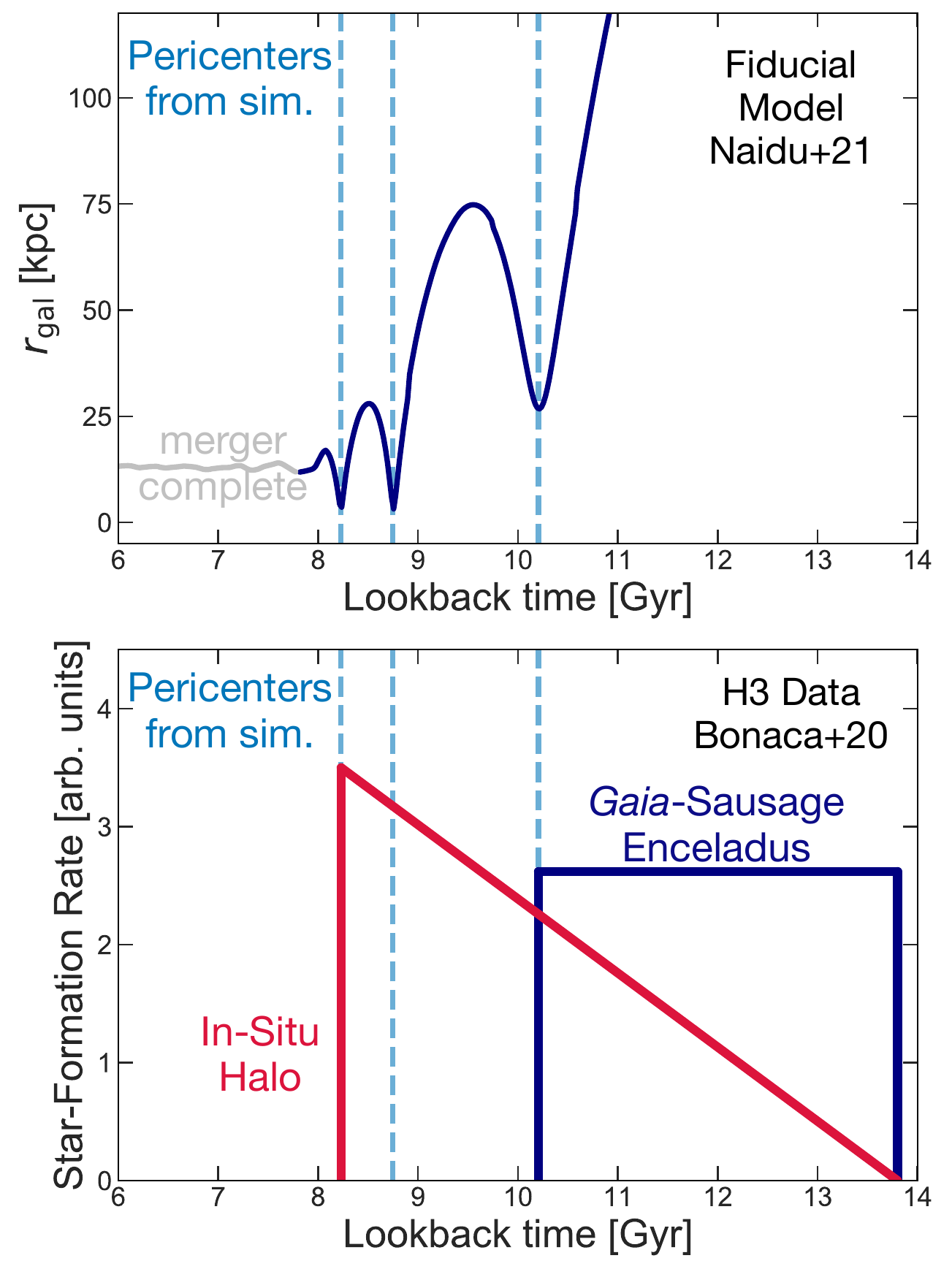}
\caption{Orbital decay profile of GSE from our fiducial simulation (\textbf{top}) compared with the \citet[][]{Bonaca20} star-formation histories (SFH) of GSE and the in-situ halo (\textbf{bottom}). The first pericenter from our simulation is assigned a lookback time coincident with the quenching of GSE in the data (10.2 Gyrs). The final pericenter, occurring $\approx2$ Gyrs later, lines up remarkably well with the truncation of the SFH of the in-situ halo, suggesting a causal relationship. After this pericenter, what is left of GSE is no longer massive/dense enough to kick stars out of the disk into the in-situ halo (bottom-center panel, Figure \ref{fig:fiducial_model}).}
\label{fig:sfh_annotate}
\end{figure}

By measuring the star-formation histories (SFH) of GSE and the in-situ halo with precise ($10\%$ median uncertainty) ages of MSTO stars, \citet[][]{Bonaca20} uncovered a 2 Gyr offset between the quenching of GSE at $\approx10$ Gyrs, and the age of the youngest stars in the in-situ halo ($\approx8$ Gyrs). In Figure \ref{fig:sfh_annotate} we compare the GSE orbital decay profile from our fiducial model with the observed SFHs --  a unified picture that accounts for the 2 Gyr offset emerges. In particular, the offset is the gap between the first and final pericentric passages.

While GSE did not lose too many stars at its first pericentric passage ($\approx25$ kpc, bottom-center panel of Fig. \ref{fig:fiducial_model}), it likely lost a good fraction of its gas. This first pericentric passage is when the SFH of GSE abruptly declines ($\approx10$ Gyrs ago, and at 0.75 Gyrs in the simulation). The final pericentric passage occurs exactly 2 Gyrs later ($\approx8$ Gyrs ago, 2.75 Gyrs in the simulation), and this is when the youngest stars in the in-situ halo are kicked out of the disk. This timeline also accounts for why the in-situ halo contains a negligible fraction of low-$\alpha$ stars, since at ${>}8$ Gyrs the high-$\alpha$ sequence was the dominant component of the disk \citep[e.g.,][]{Lian20}. Note that the second and third pericenters in the fiducial model occur in rapid succession, separated by only $\approx0.5$ Gyrs -- it is thus also possible that the second pericenter produced the bulk of the in-situ halo but the ages are not yet precise enough to be conclusive. The larger point is that there is a $\approx1.5-2$ Gyr lag between first pericenter and the deeper plunging orbit through the disk at the second and final pericenters, and this time lag agrees well with the age difference between GSE and the in-situ halo.

Another tantalizing aspect of the \citet[][]{Bonaca20} in-situ halo SFH is that it is not entirely smooth, and shows 3-4 sharp, bursty spikes just after the GSE SFH begins declining. These might be genuine starbursts sparked in the early disk by GSE, as expected from simulations \citep[e.g.,][]{Bignone19} and seen in the case of Sagittarius' predicted orbit crossing the disk  \citep[e.g.,][]{Lian20,Ruiz-Lara20}. Larger samples of stars with precise ages in the in-situ halo will help confirm these bursts, pinpoint their exact timing, and thus provide a completely independent test of our proposed GSE orbit.

\subsection{Net rotation of GSE}

\begin{figure}
\centering
\includegraphics[width=\linewidth]{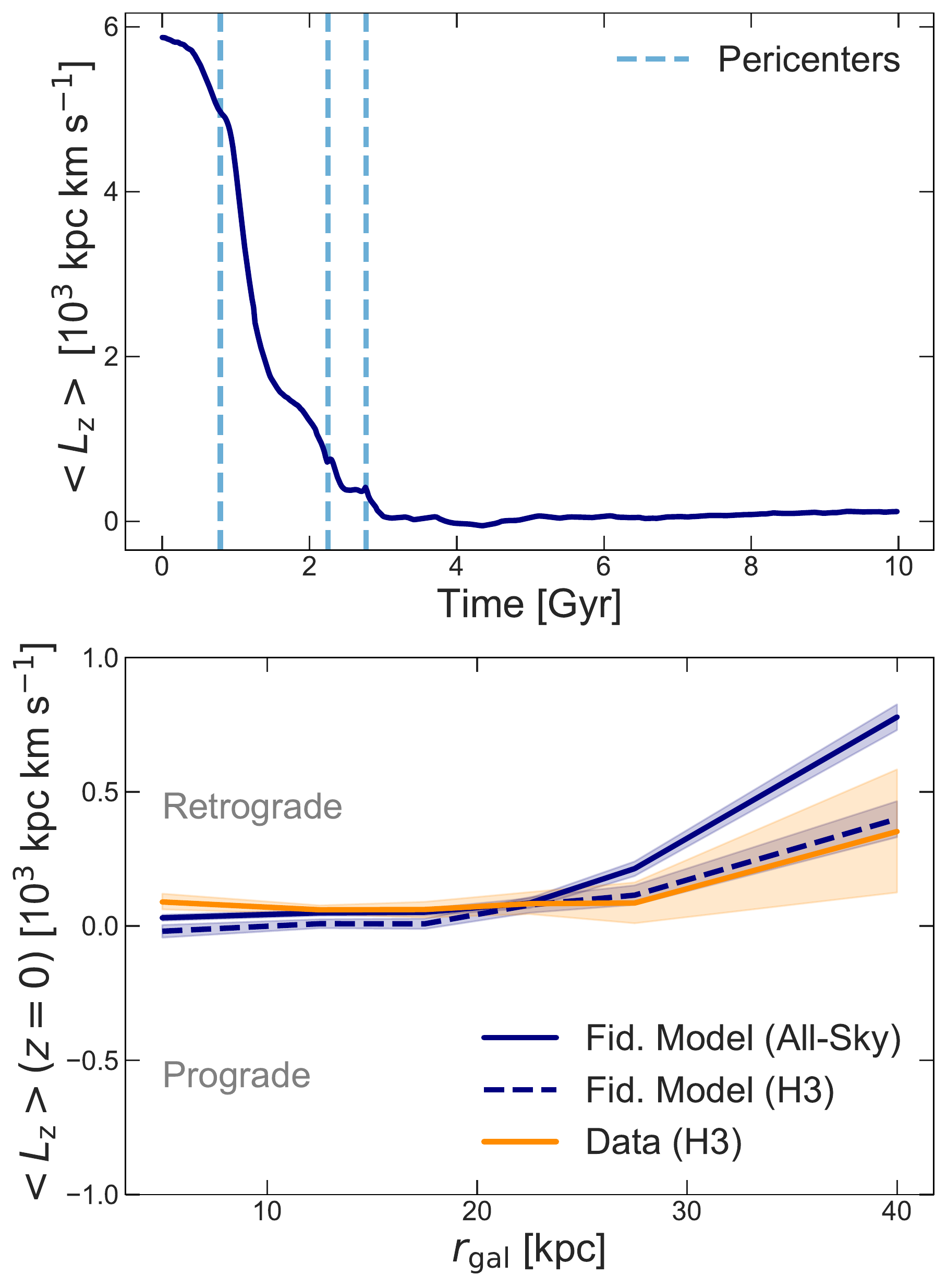}
\caption{\textbf{Top:} Evolution of the GSE mean angular momentum, $\langle L_{\rm{z}}\rangle$, in the fiducial simulation. The mean momentum is highly retrograde at infall, but following radialization by dynamical friction, the debris today has $\langle L_{\rm{z}}\rangle$ of $\approx170$ kpc km s$^{-1}$. \textbf{Bottom:} Present-day $\langle L_{\rm{z}}\rangle$ as a function of distance. $\langle L_{\rm{z}}\rangle$ is very weakly retrograde within 25 kpc ($<100$ kpc km s$^{-1}$). At larger distances, as the fraction of debris stripped early in the merger increases, the net rotation grows increasingly retrograde. Interestingly, $\langle L_{\rm{z}}\rangle$ averaged over the sky (solid navy) is more retrograde than within the H3 footprint (dashed navy) -- this can be understood via Figure \ref{fig:allsky}, where we see the stream-like, highly retrograde debris at $|l|>90^{\circ}$ occurs outside the survey footprint.}
\label{fig:net_rot}
\end{figure}

\label{sec:netrot}

The existence of Arjuna provides compelling evidence that GSE entered the MW on a highly retrograde orbit, even though the bulk of its present-day debris is radial. We demonstrate this in Figure \ref{fig:net_rot}, where in the top panel we plot the evolution of $\langle L_{\rm{z}}\rangle$ for GSE stars with time. GSE has an initial $\langle L_{\rm{z}}\rangle\approx6000$ kpc km s$^{-1}$, but in a few Gyrs it is radialized to $\langle L_{\rm{z}}\rangle\approx0$. 

In the bottom panel of Figure \ref{fig:net_rot} we plot $\langle L_{\rm{z}}\rangle$ for the GSE debris as a function of $r_{\rm{gal}}$. While there is very little mean rotation within 25 kpc, at larger distances GSE debris grows increasingly retrograde, reaching $\approx750$ kpc km s$^{-1}$ by $r_{\rm{gal}}\approx30-50$ kpc. This increase corresponds to a larger fraction of stars that were stripped early in the merger, when the bulk motion of GSE was still retrograde. Interestingly, the all-sky net rotation is higher than in the H3 sample by a factor of $\approx2\times$. This can be understood via the bottom panels of Figure \ref{fig:allsky} that depict all-sky maps of the GSE debris at $r_{\rm{gal}}>30$ kpc. The highly retrograde ``arms" of debris at $|l|>90^{\circ}$ lie at $|b|<40^{\circ}$, resulting in a less retrograde $\langle L_{\rm{z}}\rangle$ within the H3 footprint.

The transition between radial to retrograde rotation in the bottom panel of Figure \ref{fig:net_rot} may remind readers of the ``dual halo" scenario \citep{Carollo07,Carollo10,Beers12}. These authors integrated orbits of local halo samples ($d_{\rm{helio}}<4$ kpc) to infer the halo was comprised of an ``inner halo" ($r_{\rm{gal}}\lesssim15$ kpc, [Fe/H]$=-1.6$, small net prograde rotation) and an ``outer halo" ($r_{\rm{gal}}\sim20-50$ kpc, [Fe/H]$=-2.2$, mean retrograde rotation). Updating this analysis with \textit{Gaia}, \citet[][]{Carollo20} observe that GSE stars and other disk populations may constitute the inner halo, while the outer halo is composed of a variety of retrograde structures. The radial trend in GSE $\langle L_{\rm{z}} \rangle$ seen in our data and simulation is in qualitative agreement with the dual halo scenario. However, the metallicity of GSE ($\langle\rm{[Fe/H]}\rangle=-1.15$) and its flat gradient (see \S\ref{sec:fehgrad}) do not fit neatly with either the \citet[][]{Carollo10} inner or outer halo.  Further work is needed to understand the relationship between local halo samples and the global halo. For now it is clear that directly surveying the distant halo is critical to recovering the highly retrograde debris of GSE, since it is much more prominent beyond the solar circle (the $L_{\rm{z}}>1.5$ fraction at $d_{\rm{helio}}<5$ kpc is $\approx10\times$ lower than at $d_{\rm{helio}}>30$ kpc).

\begin{figure*}
\centering
\includegraphics[width=0.9\linewidth]{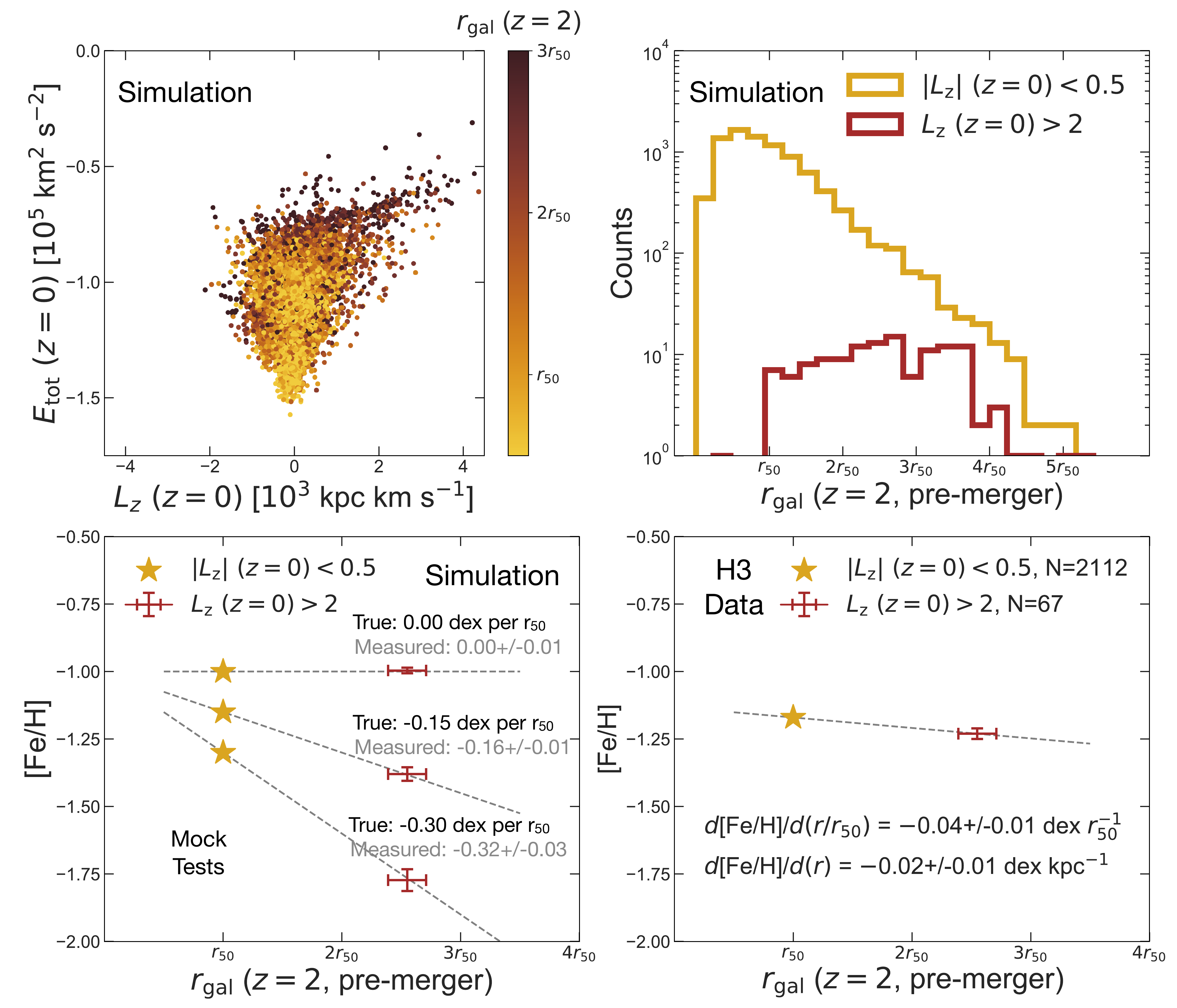}
\caption{Inferring the [Fe/H] gradient of GSE. \textbf{Top Left:} $E-L_{\rm{z}}$ diagram of GSE debris from the fiducial simulation, with stars colored by their location in the pre-merger GSE disk. Highly retrograde, high energy stars preferentially arise from the outer disk. \textbf{Top Right:} Radial distribution of GSE debris in the pre-merger disk ($r_{\rm{gal}} (z=2)$) in two bins of present-day angular momenta ($L_{\rm{z}} (z=0)$). These distributions quantify the trend seen in the left panel and motivate a mapping between $r_{\rm{gal}} (z=2)$ and $L_{\rm{z}} (z=0)$. \textbf{Bottom Left:} We exploit the trend between $L_{\rm{z}} (z=0)$ and $r_{\rm{gal}} (z=2)$ to measure [Fe/H] gradients across the GSE disk. Radial debris ($|L_{\rm{z}}|<0.5$) that traces $r_{\rm{50}}$ is shown as a golden star, whereas highly retrograde debris that traces $2.5\times r_{\rm{50}}$ ($|L_{\rm{z}}|<0.5$) is shown in brown. The radial debris is sampled by $>2000$ stars so the error on the mean is smaller than a point on this plot while the retrograde debris is sampled with 67 stars. The mock tests show excellent recovery of the true gradient (see \S\ref{sec:fehgradient}). \textbf{Bottom Right:} The [Fe/H] gradient in GSE is measured between the highly retrograde debris which preferentially arises from the outer disk and radial debris which largely arises from the inner disk. The inferred gradient is fairly shallow and is interpreted in \S\ref{sec:gradientdiscuss}.} 
\label{fig:fehgrad}
\end{figure*}

\subsection{Sequoia \& I'itoi as disrupted satellites of GSE}
\label{sec:sequoia}

Apart from Arjuna, Sequoia and I'itoi are the other prominent structures in the high-energy retrograde halo. Chemical analyses show Sequoia to have an [Fe/H]$\approx-1.6$, exactly as seen in Figure \ref{fig:data}, and that it has a ``knee" characteristic of dwarf galaxies, distinct from the GSE knee, in the [Fe/H] vs. [$\alpha$/Fe] plane \citep[e.g.,][]{Matsuno19,Monty19,Aguado20}. Nonetheless, it has been argued that Sequoia might not be a dwarf galaxy at all, and that it may in fact be debris from the outer regions of GSE, which in this scenario has a steep metallicity gradient \citep[][]{Koppelman19HS,Koppelman20b,Helmi20}. 

We offer an alternative scenario, wherein GSE has a rather flat metallicity gradient. This is supported by Arjuna's [Fe/H] that arises from the outer disk being similar to the radial debris' [Fe/H] that arises from the inner disk (Figure \ref{fig:arjuna_origin}).  If GSE has a stellar mass of $5\times10^{8} M_{\odot}$ as our numerical experiments and literature constraints suggest, then the stellar mass of Sequoia must be $\approx10^{7} M_{\odot}$ as per the relative star counts of these two structures ($<1/42$) in \citet[][]{Naidu20}. Note that this is $5-10\times$ less massive than estimated in \citet[][]{Myeong19} and \citet[][]{Kruijssen20}.  The low mass and highly retrograde phase-space position of Sequoia are consistent with it being a satellite of GSE ($\approx1:10$ by total mass according to the \citealt[][]{Behroozi19} stellar mass halo mass relation). As we have shown in this work, stars from the outer regions of GSE end up preferentially on high-energy, retrograde orbits (e.g., Figure \ref{fig:fehgrad}) -- one would expect this trend to also hold for a satellite stripped from the outer regions of GSE. This argument also applies to the more metal-poor I'itoi in Figure \ref{fig:data}, which is chemically distinct and consistent with being a dwarf, while showing integrals of motion indistinguishable from those of Arjuna and Sequoia. Detailed satellite-of-satellite simulations and ages for Sequoia \& I'itoi stars are important to test this scenario.

\section{The metallicity gradient of GSE}
\label{sec:fehgrad}

In the previous section we discussed the properties of a particular simulation that matches a variety of observational constraints. A key feature of this simulation is that the retrograde debris was stripped first and was on average at greater distances within the progenitor system than the radial debris.  In this section we exploit this property to infer the metallicity {\it gradient} within the progenitor system.

\subsection{Method \& measurement}
\label{sec:fehgradient}

In Figure \ref{fig:fehgrad} we quantify the radial distribution of stars in the pre-merger disk as a function of present-day angular momenta in our fiducial model. We find the $|L_{\rm{z}}|<0.5$ stars within the H3 footprint arise from a mean radius of $1.00^{+0.02}_{-0.02}\times r_{\rm{50}}$ in the GSE disk whereas the $L_{\rm{z}}>2$ stars arise from a mean radius of $2.55^{+0.16}_{-0.15}\times r_{\rm{50}}$. The choice to compare $L_{\rm{z}}>2$ stars with $L_{\rm{z}}<0.5$ stars is to maximize the contrast in the average pre-merger disk location.

We perform mock tests where we inject flat (0 dex per $r_{\rm{50}}$), moderate ($-0.1$ dex per $r_{\rm{50}}$), and steep ($-0.3$ dex per $r_{\rm{50}}$) metallicity gradients into our model and then attempt to recover them with an H3-like survey. We assume 0.05 dex uncertainties on individual measurements of [Fe/H] (the median uncertainty of our sample), a 10$\%$ distance error while computing $L_{\rm{z}}$, and draw the exact number of stars as in the data (2112 at $|L_{\rm{z}}|<0.5$ and 67 at $L_{\rm{z}}>2$). In all cases we were able to recover the true metallicity gradient across the disk by comparing the observed $|L_{\rm{z}}|<0.5$ and $L_{\rm{z}}>2$ stars within $10\%$ (bottom-left panel, Figure \ref{fig:fehgrad}). In the regime of very steep, unphysical gradients (e.g., -1 dex $r_{\rm{50}}^{-1}$) the method overestimates the steepness of the metallicity gradient by $\approx10\%$ because the averages no longer capture the rapid changes across the disk. A hint of this is seen in the $-0.3$ dex per $r_{\rm{50}}$ curve in Figure \ref{fig:fehgrad}. Nonetheless, for the physical regimes of interest, these tests show that angular momenta act as a superb proxy for the average location of stars in the pre-merger disk.

From Figure \ref{fig:data} it is already apparent that the metallicity gradient between the inner and outer disk populations, i.e. $|L_{\rm{z}}|<0.5$ ($\approx 1{\times}r_{\rm{50}}$)  and $|L_{\rm{z}}|>2$ ($\approx 2.5{\times}r_{\rm{50}}$) stars must be weak. The bootstrapped mean metallicity of the $|L_{\rm{z}}|<0.5$ GSE debris is [Fe/H]$=-1.17^{+0.01}_{-0.01}$ and that of the $|L_{\rm{z}}|>2$ GSE+Arjuna derbis is [Fe/H]$=-1.23^{+0.02}_{-0.02}$. This translates to a weak metallicity gradient of $-0.04\pm0.01$ dex $r_{\rm{50}}^{-1}$, comparable to Fornax ($-0.02\pm0.02$ dex $r_{\rm{50}}^{-1}$) and Ursa Minor ($-0.06\pm0.05$ dex $r_{\rm{50}}^{-1}$), which have the weakest [Fe/H] gradients of the \citet[][]{Kirby11} dwarfs. The corresponding [$\alpha$/Fe] gradient, also very weak, is $+0.02\pm0.01$ dex $r_{\rm{50}}^{-1}$.

Our definition for Arjuna truncates its MDF at [Fe/H]$<-1.5$ to avoid contamination from Sequoia, which introduces a bias in the mean [Fe/H]. Simultaneously, we also need to account for Sequoia stars at [Fe/H]$>-1.5$ that bias us to lower [Fe/H]. By fitting two Gaussians to the Sequoia and Arjuna MDFs in Figure \ref{fig:data} we find that these effects cancel out and the mean metallicity of the GSE+Arjuna $L_{\rm{z}}>2$ debris shifts imperceptibly within our stated errors ([Fe/H]$=-1.22^{+0.02}_{-0.02}$). Also note that the translation between $L_{\rm{z}}$ and $r_{\rm{50}}$ is entirely dependent on our merger model -- our error bars do not reflect the systematic uncertainty that our model may not be the only model that fits the constraints. 

\subsection{Broader context: reconstructing the stellar metallicity gradient of a $z\approx2$ galaxy}
\label{sec:gradientdiscuss}

Radial metallicity gradients probe the interplay between star-formation, feedback, inflows, and thus provide a sensitive test for how galaxies assemble their baryons across cosmic time \citep[for recent reviews see][]{Kewley19, Maiolino19, Sanchez20}. While relatively well-studied locally, gradients at higher redshifts are challenging due to the difficulty of resolving faint objects at kpc-scales. Further, existing high-$z$ gradients are derived via nebular emission, and are thus susceptible to the many systematics inherent to measuring gas-phase metallicities (e.g., uncertainties in the ionization parameter, biases from sampling only active star-forming regions, probing only the instantaneous metallicity).

 Through our rough reconstruction of the GSE disk structure we are able to effectively access the \textit{stellar} metallicity gradient of a $z\approx2$ star-forming galaxy via its $z=0$ debris. The star-formation of GSE was abruptly truncated at $z\approx2$ \citep[][]{Bonaca20}, around its first pericenter, shortly before it was shredded (Figure \ref{fig:sfh_annotate}). GSE stars inhabiting the MW halo today retain a snapshot of their $z\approx2$ chemical state. At similar redshifts, a stellar metallicity gradient has been measured in only one other galaxy --  a highly lensed ($>10\times$), very bright ($H=17.1$, $M_{\rm{\star}}\approx 6\times10^{11} M_{\rm{\odot}}$), $z=1.98$ quiescent system \citep[][]{Jafariyazani20}. We also note that the quenched ultra-faint dwarfs retain a similar chemical record of the very early universe, though they likely probe higher redshifts (e.g., $z>6$, corresponding to the Epoch of Reionization, \citealt[][]{Brown14}) and several dex lower halo masses \citep[e.g.,][]{Simon19}.

The weak, negative metallicity gradient of GSE validates the emerging observational picture that (gas-phase) gradients are generally flat at high-$z$ and grow steeper with time \citep[e.g.,][]{Leethochawalit16b,ForsterSchreiber18,Curti20}. It is also in line with dwarf simulations that produce steep gradients only towards lower redshift due to the accumulated effect of feedback-driven puffing of centrally concentrated, ancient metal-poor populations  \citep[e.g.,][]{Mercado20,El-Badry16,El-Badry18, Ma17b}. This picture also fits well with the steep gradient inferred for the Sgr dwarf galaxy \citep[e.g.,][]{Hayes20}, which has a comparable stellar and halo mass to GSE \citep[e.g.,][]{Johnson20}, but a lower accretion redshift ($z\approx0.5$) and $\gtrapprox5$ additional Gyrs of star-formation and evolution \citep[e.g.,][]{Lian20,Ruiz-Lara20,Alfaro-Cuello19}.

Stellar metallicity gradients at the redshift and mass range studied in this work ($z\approx2$, $M_{\star}\approx5\times10^{8} M_{\rm{\odot}}$) will be generally inaccessible even to \textit{JWST} and upcoming ELTs (Extremely Large Telescopes), underscoring the immense promise of ``near-field galaxy evolution" as a complementary route to studying the high-$z$ universe through halo debris \citep[e.g.,][]{Boylan-Kolchin15,Boylan-Kolchin16,Weisz14}. GSE stars are beginning to be used in ``$z\approx2$" studies to understand the chemistry of the early Universe \citep{Molaro20,Simpson20,Matsuno21} -- our simulations will add rich context to these works (e.g., by mapping the phase-space of GSE stars to their pre-merger disk location and the time they were stripped). Similar reconstructions will soon be possible for other less massive dwarfs accreted at a variety of redshifts as our census of halo substructure grows more complete.

\section{Summary}
\label{sec:summary}

We have used the H3 Survey to study the $z\approx2$ GSE merger. Our unique sample of $\approx2800$ GSE stars has full 6D phase-space data, abundances ([Fe/H], [$\alpha$/Fe]), is unbiased in metallicity, and encompasses the farthest reaches of GSE debris. We systematically explore a large grid ($\approx500$) of high resolution ($10^{5} M_{\rm{\odot}}$) N-body simulations to reproduce the H3 constraints (summarized in \S\ref{sec:constraints}), and re-simulate the most promising configurations ($N\approx20$) at an even higher resolution ($10^{4} M_{\rm{\odot}}$). Our grid spans a plausible range of physical (size, mass) and orbital (circularity, inclination, disk spin, sense of orbit) parameters (Tables \ref{table:gestruc}, \ref{table:mwstruc}, \ref{table:orbitalparams}). We find the merger and its resultant debris have the following characteristics:

\begin{itemize}
    \item GSE arrived in the Galaxy on a highly retrograde orbit. The Arjuna structure identified in \citet[][]{Naidu20} is the retrograde debris of GSE. Despite harboring some of the most retrograde stars in the halo, the [Fe/H] and [$\alpha$/Fe] of this structure are within $\approx0.05$ dex of the radial, eccentric ($e>0.7$) GSE stars, strongly suggesting they are associated. [Figure \ref{fig:data}, \S\ref{sec:datamain}]
 
    \item A GSE of mass $M_{\star}=5\times10^{8} M_{\rm{\odot}}, M_{\rm{DM}}=2\times10^{11} M_{\rm{\odot}}$ with an $r_{\rm{50}}=2.5$ kpc that merges on a retrograde orbit with a circularity of 0.5, inclination of $15^{\circ}$, and retrograde disk spin best reproduces the H3 constraints. The GSE merger was a 2.5:1 merger -- its debris comprises $\approx20\%$ of the $z=0$ Milky Way dark matter halo and $\approx50\%$ of its stellar halo. [Figure \ref{fig:likelihood}, \ref{fig:fiducial_model}, \S\ref{sec:fiducial}]
    
    \item The retrograde Arjuna stars preferentially arise from the outer disk of GSE ($\approx2.5\times r_{\rm{50}}$) in our fiducial simulation. These loosely bound stars are stripped early in the merger, before GSE has been radialized by dynamical friction, and so they reflect the initial retrograde angular momentum. [Figure \ref{fig:arjuna_origin}, \S\ref{sec:arjuna_origin}]
    
    \item The net rotation of GSE at $z\approx2$ was $\langle L_{\rm{z}}\rangle\approx6000$ kpc km s$^{-1}$, but by $z=0$ it is largely radialized ($\langle L_{\rm{z}}\rangle\approx170$ kpc km s$^{-1}$). As a function of distance, rotation of GSE debris is weak within 25 kpc ($\langle L_{\rm{z}}\rangle<100$ kpc km s$^{-1}$), but grows to $\langle L_{\rm{z}}\rangle\approx750$ kpc km s$^{-1}$ at $r_{\rm{gal}}\approx30-50$ kpc as the fraction of debris stripped early in the merger grows with distance. [Figure \ref{fig:net_rot}, \S\ref{sec:netrot}]
\end{itemize}

Even though our fiducial simulation is selected purely based on the H3 Survey $r_{\rm{gal}}$ and $L_{\rm{z}}$ distributions, it self-consistently reproduces and explains the following phenomena:

\begin{itemize}

    \item The shape of the inner halo ($<30$ \rm{kpc}) is set by GSE, which is its most dominant constituent. GSE debris defines an elongated triaxial ellipsoid with axes ratios $10:7.9:4.5$, in remarkable agreement with \textit{Gaia} RR Lyrae constraints \citep[][]{Iorio19}. The major axis of the debris is at $\approx35^{\circ}$ to the disk plane, and at $\approx45^{\circ}$ to the Galactic X/Y direction. The orientation tracks the second and final apocenters that occur on either side of the plane. GSE loses most of its stars between these apocenters. The tilted triaxial halo is a significant departure from planar, prolate models typically used to model the Galactic stellar and dark matter halos. [Figure \ref{fig:shape}, \ref{fig:allsky}, \S\ref{sec:shape}]
    
    \item The Hercules-Aquila Cloud (HAC) and Virgo Overdensity (VOD) occur on either end of the triaxial ellipsoid's major axis and emerge due to apocenter pile-up of GSE debris. The HAC and VOD are spatially proximal to the penultimate and final apocenter of the GSE orbit respectively.  [Figure \ref{fig:HACVOD}]
    
    \item The $\approx2$ Gyr gap between the quenching of GSE and the cessation of star-formation in the in-situ halo \citep[][]{Bonaca20} is precisely the gap between the first and final pericentric passages. At first pericenter the star formation within GSE is truncated, and by final pericenter it is no longer massive enough to kick stars out of the primordial disk into the in-situ halo. [Figure \ref{fig:sfh_annotate}, \S\ref{sec:timing}]
    
\end{itemize}

We make the following predictions based on our fiducial simulation:

\begin{itemize}

    \item The inner halo has a ``double-break" density profile. The two breaks (at $\approx15$ kpc, $\approx30$ kpc) correspond to the second and final apocenters and are described by the following power-law ($\rho \propto r_{\rm{gal}}^{\rm{\alpha}}$) coefficients: $\alpha\ ({<}15\ \rm{kpc})=-1.1$, $\alpha\ (15-30\ \rm{kpc})=-3.3$. The profile shows strong variations across the sky both in the normalization ($\approx5-10\times$) as well as the shape ($\Delta\alpha\approx1.5$). Interestingly, several studies have fit single-break profiles for the inner halo but disagree about the break location, with some finding a break at $\approx15-20$ kpc and others at $\approx25-30$ kpc. Our proposed profile may resolve this tension. [Figure \ref{fig:profile}, \S\ref{sec:profile}]
    
    \item The outer halo ($r_{\rm{gal}}>30$ kpc) contains $\approx10\%$ of the GSE stellar mass. This debris manifests as highly retrograde, stream-like structures that await discovery.  Approximately $50\%$ of this debris lies within $20^{\circ}$ of the Sgr orbital plane. [Figure \ref{fig:allsky}, \ref{fig:allsky2}, \S\ref{sec:outerhalo}]

    \item The Sequoia and I'itoi dwarfs, which have integrals of motion virtually indistinguishable from Arjuna, may also have been stripped from the outer regions of GSE. These systems may have once constituted a group like the Magellanic Clouds. [\S\ref{sec:sequoia}]
    
\end{itemize}

Finally, we use our fiducial simulation to reconstruct the stellar metallicity gradient in a $z\approx2$ star-forming galaxy (GSE):

\begin{itemize}

    \item Radial GSE debris ($|L_{\rm{z}}|<0.5$) originates from the inner disk ($\approx r_{\rm{50}}$) while retrograde debris ($L_{\rm{z}}>2$) arises from the outer disk ($\approx2.5\times r_{\rm{50}}$). Capitalizing on this trend, we measure a stellar metallicity gradient of $-0.04\pm0.01$ dex $r_{\rm{50}}^{-1}$ and [$\alpha$/Fe] gradient of $+0.02\pm0.01$ dex $r_{\rm{50}}^{-1}$. Stellar abundance gradients for star-forming galaxies at $z\approx2$ will be inaccessible even to \textit{JWST} -- our measurement underscores the immense promise of ``near-field galaxy evolution" with halo debris as a complementary route to the high-$z$ universe. [Figure \ref{fig:fehgrad}, \S\ref{sec:fehgradient}]
    
\end{itemize}

We once again emphasize that our fiducial simulation is \textit{a} possible configuration for the GSE merger and not necessarily \textit{the} configuration. However, it is quite successful at not only replicating the H3 data, but also reproducing and explaining disparate phenomena across the Galaxy. Further, it makes specific, verifiable predictions that can be tested with existing and upcoming datasets. We foresee this model being used to drive progress on multiple fronts. For instance, the detailed phase-space distribution of the enormous amount of GSE DM left unexplored in this work could prove critical to designing DM detection experiments and informing realistic models of the Milky Way potential. The in-situ halo/splashed disk can be developed into a sensitive probe of the physical \& chemical structure of the primordial MW disk. Taken together, GSE ($2\times10^{11} M_{\rm{\odot}}$, this work), Sgr ($\approx1\times10^{11} M_{\rm{\odot}}$, e.g., \citealt[][]{Johnson20}), and the LMC ($\approx1.3\times10^{11} M_{\rm{\odot}}$, e.g., \citealt[][]{Erkal19}) account for almost the entirety of the growth of the MW since $z\approx2$ -- the census of the MW's significant mergers in the last 10 Gyrs is now complete.

\facilities{MMT (Hectochelle), \textit{Gaia}}

\software{
    \package{IPython} \citep{ipython},
    \package{matplotlib} \citep{matplotlib},
    \package{cmasher} \citep{cmasher},
    \package{numpy} \citep{numpy},
    \package{scipy} \citep{scipy},
    \package{jupyter} \citep{jupyter},
    \package{gala} \citep{gala1, gala2},
    \package{Astropy}
    \citep{astropy1, astropy2},
    \package{Gadget-2,3,4} \citep{Springel05,Springel08,Springel20},
    \package{GalIC} \citep{Yurin14},
    \package{Glue} \citep{glue1,glue2}
    }
    
\acknowledgments{We thank Volker Springel for sharing \texttt{GADGET-3} with us, and for making \texttt{GADGET-2} and \texttt{GADGET-4} publicly available. We are grateful to Nico Garavito-Camargo, Harshil Kamdar, and Gus Beane for advice on simulations. We acknowledge helpful feedback on AB's April 2020 colloquium at the University of Cambridge, useful suggestions from Vasily Belokurov, Azadeh Fattahi, and Jorge Pe{\~n}arrubia, and an illuminating discussion with the Cambridge Streams Group. RPN gratefully acknowledges an Ashford Fellowship granted by Harvard University. CC acknowledges funding from the Packard foundation. YST is supported by the NASA Hubble Fellowship grant HST-HF2-51425.001 awarded by the Space Telescope Science Institute. We thank the Hectochelle operators Chun Ly, ShiAnne Kattner, Perry Berlind, and Mike Calkins, and the CfA and U. Arizona TACs for their continued support of the H3 Survey. 

This paper uses data products produced by the OIR Telescope Data Center, supported by the Smithsonian Astrophysical Observatory. The computations in this paper were run on the FASRC Cannon cluster supported by the FAS Division of Science Research Computing Group at Harvard University. This work has made use of data from the European Space Agency (ESA) mission
{\it Gaia} (\url{https://www.cosmos.esa.int/gaia}), processed by the {\it Gaia}
Data Processing and Analysis Consortium (DPAC,
\url{https://www.cosmos.esa.int/web/gaia/dpac/consortium}) \citep{dr2ack1, dr2ack2}. Funding for the DPAC
has been provided by national institutions, in particular the institutions
participating in the {\it Gaia} Multilateral Agreement.}
    
\bibliography{MasterBiblio}
\bibliographystyle{apj}

\end{CJK*}
\end{document}